\documentclass[pra,aps,showpacs,showkeys,twocolumn,twoside]{revtex4}

\usepackage{amsmath}
\usepackage{amssymb}
\usepackage{amscd}

\usepackage{latexsym}
\usepackage{epsfig}
\usepackage{graphics}

\newtheorem{defi}{Definition}
\newtheorem{lemma}[defi]{Lemma}
\newtheorem{thm}[defi]{Theorem}

\newtheorem{rem}[defi]{Remark}

\newtheorem{expl}[defi]{Example}

\newcommand{\qed}{\hfill $\Box$}
\newcommand{\tr}{{\operatorname{Tr}\,}}
\newcommand{\id}{{\operatorname{id}}}
\newcommand{\supp}{{\operatorname{supp}\,}}
\newcommand{\rank}{{\operatorname{rank}\,}}
\newcommand{\conv}{{\operatorname{conv}}}
\newcommand{\bra}[1]{{\langle{#1}|}}
\newcommand{\ket}[1]{{|{#1}\rangle}}
\newcommand{\ketbra}[1]{{\ket{#1}\!\bra{#1}}}
\newcommand{\C}{{\mathbb{C}}}
\newcommand{\R}{{\mathbb{R}}}

\newcommand{\fset}[1]{{\mathcal{#1}}}
\newcommand{\Rho}{{\sf P}}
\newcommand{\Nu}{{\sf N}}

\newcommand{\E}{{\mathbb{E}}}
\newcommand{\1}{{\openone}}

\newlength{\blank}
\newlength{\equalsign}
\newenvironment{beweis}[1][{\hspace{-\blank}}]{{\noindent\emph{Proof~{#1}.\ }}}{\hfill $\Box$\vskip 0.5\baselineskip}
\newlength{\lueckeeins}                          
\newlength{\lueckezwei}                          
\newlength{\lueckedrei}                          
\begin{document}


\title{``Extrinsic'' and ``intrinsic'' data in quantum measurements: \protect\\
asymptotic convex decomposition of positive operator valued measures}
\author{Andreas Winter}
\email{winter@cs.bris.ac.uk}
\affiliation{Department of Computer Science, University of Bristol,\\
Merchant Venturers Building, Woodland Road, Bristol BS8 1UB, United Kingdom}
\date{19th November, 2002}


\begin{abstract}
  We study the problem of separating the data produced by a given quantum
  measurement (on states from a memoryless source which is
  unknown except for its average state),
  described by a positive operator valued measure (POVM),
  into a ``meaningful'' (\emph{intrinsic}) and a ``not meaningful''
  (\emph{extrinsic}) part.
  \par
  We are able to give an asymptotically tight separation of this
  form, with the ``intrinsic'' data quantfied by the Holevo
  mutual information of a certain state ensemble associated 
  to the POVM and the source, in a model that can be viewed
  as the asymptotic version of the convex decomposition of POVMs
  into extremal ones.
  This result is applied to a similar separation therorem
  for quantum instruments and quantum operations, in their
  Kraus form.
  \par
  Finally we comment on links to related subjects: we
  stress the difference between data and information (in
  particular by pointing out that \emph{information} typically
  is strictly \emph{less} than \emph{data}), derive
  the Holevo bound from our main result, and look at its
  classical case: we show that this includes the solution to the problem
  of extrinsic/intrinsic data separation with a \emph{known} source,
  then compare with the well--known notion of sufficient statistics.
  The result on decomposition of quantum operations is used to
  exhibit a new aspect of the concept of entropy exchange of an open dynamics.
  \par
  An appendix collects several estimates for mixed state fidelity and
  trace norm distance, that seem to be new, in particular a construction
  of \emph{canonical purification} of mixed states that turns out to be
  valuable to analyze their fidelity.
\end{abstract}



\maketitle

%
%
%

\section{The problem}
\label{sec:introduction}
Consider a quantum system, represented by a Hilbert space ${\cal H}$
(which we assume to be of dimension $d<\infty$ in the sequel), and
a measurement on this system, described by a positive operator
valued measure (POVM) ${\bf a}=(a_1,\ldots,a_m)$,
$a_j\in{\cal B}({\cal H})$ such that $a_j\geq 0$ and
$\sum_j a_j=\1$.
\par
Following~\cite{massar:popescu} and~\cite{winter:massar:POVMcompr} we
shall be concerned with the question ``How much information is obtained
by ${\bf a}$?'', beginning with a clarification what this question
should mean at all. Imagine that a family of states (represented
by density operators) $\rho_i$ on ${\cal H}$ is given, let us
say with a priori probabilities $p_i$, such that the density operator
of this source of states is $\rho=\sum_i p_i\rho_i$, then
the ``information'' in question could mean the information in $j$
about $i$, and one way to quantify it would be given by Shannon's
mutual information~\cite{shannon:info} $I(i\wedge j)$.
Note that this is in general less than the amount of raw data,
which is operationally quanified by the entropy of the distribution
of the $j$: $H(\lambda)$, with $\lambda_j=\tr(\rho a_j)$,
due to Shannon's source coding theorem~\cite{shannon:info}.
\par
This choice however is rather arbitrary: asking about the identity
of the state from a list. Why not allowing a different list, or asking
for some property of the state. Also, mutual information is a
measure of correct identification; but what if we need only
``almost correct'' identification, as in quantum statistical
detection theory~\cite{helstrom:detection}?
\par
It seems hence that specifying \emph{the information} in measurement
results, or even only the \emph{amount}, in an operationally satisfying
way, is problematic, and one reason might be the complementarity
of quantum mechanics: qualitatively, accessing some observable property
optimally entails rather poor performance for others.
Nevertheless, it is quite obvious intuitively that in almost any
POVM there is ``quantum noise'', i.e. redundancy put into the $j$
by the very quantum mechanical probability rule, most simply due
to nonorthogonality of the operators $a_j$, for example in an
overcomplete system (see e.g.~\cite{holevo:statistical}).
\par
Our approach will thus be from the opposite end: instead of attempting
the impossible, defining what ``useful'' means in any circumstances,
we adopt a very simple criterion of use\emph{less}ness: statistical
independence form the measured states, because independent randomness
can be generated from outside without accessing the quantum system.
On the other hand we do not permit a distortion of the measurement itself,
so that we are forced to consider a \emph{simulation} of the original
measurement by means of, first, a random choice $\nu$ of a measurement
${\bf a}^{(\nu)}$ from
a list and, second, computation of a result from the outcome of this
measurement and the random choice, such that the statistical distribution
of these results is indistinguishable from the ones of the original
measurements, on any prepared state.
\par
Because we can absorb the computation of the results into the labelling
of the ${\bf a}^{(\nu)}$, this means that we aim at finding such POVMs,
whose indices are labelled by the same $j$ as ${\bf a}$ and probabilities
$x_\nu$, such that
\begin{equation}
  \label{eq:convex:decomp}
  {\bf a}=\sum_\nu x_\nu{\bf a}^{(\nu)},\text{ i.e. }
                \forall j\ a_j=\sum_\nu x_\nu a^{(\nu)}_j.
\end{equation}
(The operators must be the same because otherwise there would states
that induce distinguishable outcome distributions. Below we will introduce
an element of approximation into this scheme).
\par
Why should we want to do such a decomposition, interesting though the
structure exhibited (convex set of POVMs) might be mathematically?
Observe that each ${\bf a}^{(\nu)}$ has its distribution of outcomes,
with the probabilities
$\lambda^{(\nu)}_j=\tr\left(\rho a^{(\nu)}_j\right)$
of $j$ conditional on $\nu$. Shannon's source coding theorem~\cite{shannon:info}
quantifies the amount of data in such a source as the (Shannon) entropy
$$H\bigl(\lambda^{(\nu)}\bigr) = \sum_j -\lambda^{(\nu)}_j\log\lambda^{(\nu)}_j,$$
by compression (we note that in this paper all $\log$s and $\exp$s are to basis $2$).
Hence, on average, one needs 
$$H(j|\nu):=\sum_\nu x_\nu H\left(\lambda^{(\nu)}\right)$$
bits to faithfully compress the data ($j$), given $\nu$ as side--information.
\par
This motivates the study of the function
\begin{equation}
  \label{eq:deltafunction}
  \delta(\rho,{\bf a}):=\min \left\{H(j|\nu):{\bf a}=\sum_\nu x_\nu {\bf a}^{(\nu)}\right\},
\end{equation}
which is the minimum data rate (in Shannon's sense) for
\emph{exact} reconstruction of the data.

\begin{expl}
  Look at a qubit system, $\C^2$, with basis $\{\ket{0},\ket{1}\}$: there
  let us consider the five ``Chrysler'' states (in analogy to the
  ``Mercedes'' trine states)
  $$\ket{e_t}=\left(\cos\frac{\pi t}{5}\right)\ket{0}
                +\left(\sin\frac{\pi t}{5}\right)\ket{1},\text{ for }t=0,\ldots,4.$$
  The collection ${\bf a}=(\frac{2}{5}\ketbra{e_t})_{t=0,\ldots,4}$ is a POVM, and we can
  determine its decompositions into extremal ones: these latter are given by
  putting weights on the $\ketbra{e_t}$, and
  it is straightforward that for an extremal POVM at most $3$ can be nonzero
  (as the ``Chrysler'' states form a pentagon on the Bloch sphere equator).
  In fact, every extremal must be of the form
  $$\bigl(\alpha\ketbra{e_t},\beta\ketbra{e_{t+2}},\beta\ketbra{e_{t+3}}\bigr),
                                                                \quad t=0,\ldots,4,$$
  indices understood modulo $5$.
  From here one can determine the weights to be
  \begin{align*}
    \alpha &=1-\left(\cot\frac{2\pi}{5}\right)^2             \approx 0.5528 \\
    \beta  &=\frac{1}{2}\left(\sin\frac{2\pi}{5}\right)^{-2} \approx 0.8944
  \end{align*}
  For simplicity now look at the maximally mixed state $\rho=\frac{1}{2}\1$,
  for which it is unimportant which decomposition into these extremal POVMs
  is chosen, as all contributions $\nu$ will give the same Shannon entropy:
  \begin{equation*}\begin{split}
    \delta(\rho,{\bf a}) &= H(j|\nu)
                          = H\left(\frac{\alpha}{2},\frac{\beta}{2},\frac{\beta}{2}\right) \\
                         &= H\left(1-\beta,\frac{\beta}{2},\frac{\beta}{2}\right)          \\
                         &= H(1-\beta,\beta)+\beta \approx 1.5447
  \end{split}\end{equation*}
  In contrast, the main theorem~\ref{thm:ex-and-in} below
  will achieve a rate of $H(\rho)=1$, asymptotically.
\end{expl}
\par
The computation of $\delta(\rho,{\bf a})$ is an interesting problem in its own right
(in particular the question if anything can be gained on $\delta$ by considering
multiple copies, i.e.~the additivity problem), however we take a different approach,
bearing in mind that the operational content of Shannon's theorem involves block
coding --- i.e., a large number $l$ of independent copies of the simple system described
above, and an arbitrarily small yet nonzero error probability:
\par
Thus we are really decomposing the POVM
$${\bf a}^{\otimes l} 
           = \bigl(a_{j_1}\otimes\cdots\otimes a_{j_l}\bigr)_{j^l\in\{1,\ldots,m\}^l},$$
where we have introduced te notation $j^l=j_1\ldots j_l$ for a string of symbols,
used henceforth. And the error introduced through block compression entails that
instead of eq.~(\ref{eq:convex:decomp}) we will only have
\begin{equation}
  \label{eq:idea}
  {\bf a}^{\otimes l}\approx {\bf A}=\sum_{\nu} x_\nu {\bf A}^{(\nu)},
\end{equation}
where the $\approx$ sign is made precise to mean ``average approximation of outcome
statistics'': assuming an ensemble $\{\sigma_k,q_k\}$ with
$\sum_k q_k\sigma_k = \rho^{\otimes n}$,
there is the joint distribution of input $k$ and output $j^l$
when applying ${\bf a}^{\otimes l}$
\begin{equation}
  \label{eq:gamma}
  \gamma(k,j^l)=q_k \tr(\sigma_k a_{j^l}),
\end{equation}
and likewise for ${\bf A}$:
\begin{equation}
  \label{eq:GAMMA}
  \Gamma(k,j^l)=q_k \tr(\sigma_k A_{j^l}).
\end{equation}
Then we require that, independent of the particular ensemble,
\begin{equation}
  \frac{1}{2}\|\gamma-\Gamma\|_1=\sum_{k,j^l} \frac{1}{2}|\gamma(k,j^l)-\Gamma(k,j^l)|
               \leq \epsilon\tag{CP}.
\end{equation}
(It is not difficult to see that eq.~(\ref{eq:convex:decomp}) raised to the 
the $l^{\rm th}$ tensor power, together with Shannon compression of the outcomes
of ${\bf a}^{(\nu_1)}\otimes\cdots\otimes{\bf a}^{(\nu_l)}$ for the
\emph{probably} $\nu_1\ldots\nu_l$ yields exactly that).
Indeed we can, using the abbreviation $\omega=\rho^{\otimes l}$,
rewrite eq.~(\ref{eq:gamma}) as
$$\gamma(k,j^l)=\tr\left(\omega^{-1/2}q_k\sigma_k\omega^{-1/2}
                          \sqrt{\omega}a_{j^l}\sqrt{\omega}\right),$$
observing that the $S_k=\omega^{-1/2}q_k\sigma_k\omega^{-1/2}$ form
a POVM on ${\cal H}^{\otimes l}$ (this fact was observed before,
and used in~\cite{hjw:classification} to classify all ensembles with
a given average state). Similarly
$$\Gamma(k,j^l)=\tr\left(S_k \sqrt{\omega}A_{j^l}\sqrt{\omega}\right),$$
and we can rewrite and estimate the left hand side of (CP) as follows:
\begin{equation*}\begin{split}
  \frac{1}{2}\|\gamma-\Gamma\|_1
       &= \sum_{j^l}\sum_k \frac{1}{2}
          \left|\tr\left(S_k \sqrt{\omega}(a_{j^l}-A_{j^l})\sqrt{\omega}\right)\right| \\
       &\leq \sum_{j^l} \frac{1}{2}
          \left\| \sqrt{\omega}(a_{j^l}-A_{j^l})\sqrt{\omega} \right\|_1,
\end{split}\end{equation*}
so (CP) is in fact implied by
\begin{equation}
  \sum_{j^l} \frac{1}{2}
     \left\|\sqrt{\rho}^{\otimes l}(A_{j^l}-a_{j^l})\sqrt{\rho}^{\otimes l}\right\|_1
                                                              \leq \epsilon.\tag{CM}
\end{equation}
Notice that the condition can be phrased in a particularly
nice way introducing the quantum operations
\begin{align}
  \label{eq:POVM:op1}
  \varphi_*^{\otimes l}:\sigma &\longmapsto \sum_{j^l} \tr(\sigma a_{j^l})\ketbra{j^l}, \\
  \label{eq:POVM:op2}
  \Phi_*   :            \sigma &\longmapsto \sum_{j^l} \tr(\sigma A_{j^l})\ketbra{j^l}.
\end{align}
Namely, for a purification $\pi$ of $\rho$, (CM) is easily seen to be equivalent to
\begin{equation}
  \label{eq:POVM:purif}
  \frac{1}{2}\left\| (\id\otimes\Phi_*)(\pi^{\otimes l})
                    -\left(\id\otimes\phi_*^{\otimes l}\right)(\pi^{\otimes l})
             \right\|_1 \leq \epsilon.
\end{equation}
\par\medskip
The organization of the paper is as follows:
In section~\ref{sec:ex-and-in} we will present our main
theorem~\ref{thm:ex-and-in} and its proof, which is much more satisfying than
results in previous work~\cite{massar:popescu,winter:massar:POVMcompr},
that can now be regarded as precursors: theyy are shown to easily follow from
theorem~\ref{thm:ex-and-in} in section~\ref{sec:upgrade}.
Section~\ref{sec:strong:converse} is concerned with the
asymptotic optimality of our main theorem, a \emph{strong converse}
result, theorem~\ref{thm:strong:converse}.
After this, in section~\ref{sec:other:appl} we apply our result to a kind
of asymptotic normal form of completely positive trace preserving maps
(operations as well as instruments),
and present an extensive discussion in
section~\ref{sec:discussion}: we restate our observation
from~\cite{winter:massar:POVMcompr} that one ought to
distinguish \emph{obtained data} from \emph{information},
give a new, conceptually simple proof of the Holevo bound,
remark on the classical case of the main theorem (which
includes the problem of separating extrinsic and intrinsic
data under a \emph{known} source ensemble),
comment on the related concept of sufficient statistics,
and discuss the bearing of our results on the concept
of \emph{entropy exchange} of an open dynamics of a system.
We close with a challenging open problem.
An appendix features several not widely known facts about the mixed state
fidelity, in particular introducing \emph{canonical purifications}
of mixed states, a second appendix collects properties of
\emph{typical sequences} and \emph{typical subspaces}, used
in the main text.

\section{Separating extrinsic and\protect\\ intrinsic data}
\label{sec:ex-and-in}
We want to represent (up to a small deviation as specified by the (CM)
condition) ${\bf a}^{\otimes l}$ as a convex combination of
POVMs ${\bf A}^{(\nu)}$, with positive weights $x_\nu$, $\nu=1,\ldots,N$,
each being defined on the set $[m]^l$ and having a small number $M$
of sequences on which it is supported (i.e. where $A^{(\nu)}_{j^l}\neq 0$):
this is an even stronger requirement than the entropy condition we
had considered in the introduction.
\begin{figure}[ht]
  \includegraphics{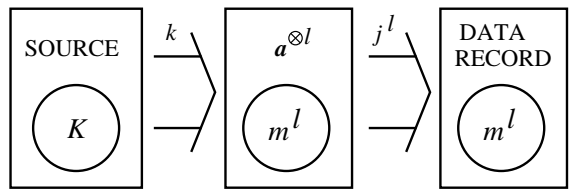}
  \caption{The source represents a number of possible states
           encountered by the POVM, but there is no way of knowing
           which is present (apart from the apriori distribution).
           The data produced by the measurement is then stored in
           a record. The rates of these processes are represented by the
           sizes of the different boxes and width of
           the data flow arrows: originally the rates of
           the source and of the measurement outcomes are both large.}
  \label{fig:data}
\end{figure}
Performing ${\bf A}$ amounts to choosing a $\nu$ (with probability
$x_\nu$), and performing
${\bf A}^{(\nu)}$, which itself can generate at most $M$ different outcomes:
the $\nu$--part of the produced data is obviously independent of the
incoming signal, while the measurement outcome (conditional on the $\nu$ chosen)
contains the useful information.
\par
Our central result is:
\begin{thm}
  \label{thm:ex-and-in}
  There exist POVMs ${\bf A}^{(\nu)}$ on $[m]^l$, $\nu=1,\ldots,N$,
  each supported on a set of cardinality at most $M$, where
  \begin{align*}
    M &=\exp \left( l I(\lambda;\hat{\rho})+O(\sqrt{l}) \right),\\
    N &=\exp \left( l \bigl(H(\lambda)-I(\lambda;\hat{\rho})\bigr)+O(\sqrt{l}) \right),
  \end{align*}
  such that for ${\bf A}=\frac{1}{N}\sum_\nu {\bf A}^{(\nu)}$
  condition (CM) is satisfied.
\end{thm}
The characteristic constant in the exponent is
$$I(\lambda;\hat{\rho})=H(\rho)-\sum_j \lambda_j H(\hat{\rho}_j),$$
the \emph{entropy defect} of the ensemble
(Lebedev and Levitin~\cite{levitin:defect}),
or the \emph{quantum mutual information}
between a sender producing letter $j$ with probability $\lambda_j$
and a receiver getting the letter state $\hat{\rho}_j$
(see~\cite{holevo:coding,schumacher:westmoreland}).
It is the difference between the von Neumann
entropy $H(\rho)=-\tr\rho\log\rho$
of the ensemble and its \emph{conditional entropy}
$H(\hat{\rho}|\lambda)=\sum_j \lambda_j H(\hat{\rho}_j)$.
\par
Observe that not only $\rho$ can be recovered from this ensemble (as its
average), but also the POVM ${\bf a}$:
$$a_j=\rho^{-1/2} \lambda_j\rho_j \rho^{-1/2}.$$
This construction is known as the
``square root measurement''~\cite{holevo:sqrt}, or
``pretty good measurement''~\cite{hausladen:pgm}.
\begin{figure}[ht]
  \includegraphics{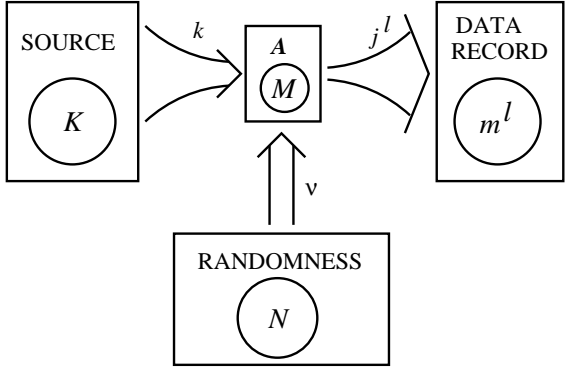}
  \caption{A nice way of picturing the content of
           theorem~\ref{thm:ex-and-in} is in the form of an
           elaborate bottleneck between source and outcomes:
           it is supplied from outside with the extrinsic
           data $\nu$, and conditional on this and the
           incoming $k$ produces the intrisic data $j^l$. Only
           the intrinsic data are correlated to the signal $k$,
           while the extrinsic data (though evidently an indispensable part
           of the whole data) is independent of it.
           To put it pointedly: while it is difficult and possibly
           ambiguous to speak of ``useful data'', one can clearly
           identify data of no import in all respects: the unrelated
           randomness $\nu$. This is put into the focus by
           theorem~\ref{thm:ex-and-in}, and our concept of usefulness
           is just the remainder after extracting as much uselessness
           as possible.}
  \label{fig:bottle}
\end{figure}
We shall give the proof of theorem~\ref{thm:ex-and-in}
in a minute, after a few preparations.
A central part of the argument is the following auxiliary
result from~\cite{ahlswede:winter:QID} that we state separately:
\begin{lemma}[Ahlswede,~Winter~\cite{ahlswede:winter:QID}, thm.\! A.19]
  \label{lemma:largedev}
  Let\\ $X_1,\ldots,X_M$ be independent identically distributed
  (i.i.d.) random variables with
  values in the algebra ${\cal L}({\cal K})$ of linear operators
  on ${\cal K}$, which are bounded between
  $0$ and $\1$. Assume that the average $\E X_\mu=\sigma\geq s\1$.
  Then for every $0<\eta<1/2$
  $$\Pr\!\left\{\frac{1}{M}\sum_{\mu=1}^M X_\mu\not\in[(1\pm\eta)\sigma]\right\}
              \!\leq 2\dim{\cal K}\exp\!\left(\! -M\frac{\eta^2 s}{2\ln 2}\right)\!,$$
  where $[(1\pm\eta)\sigma]=[(1-\eta)\sigma;(1+\eta)\sigma]$ is an interval
  in the operator order: $[A;B]=\{X\in{\cal B}({\cal K}):A\leq X\leq B\}$.
  \qed
\end{lemma}
We shall use the concepts of \emph{typical} and
\emph{conditionally typical} subspaces in the form
of~\cite{winter:ieee_strong}, which we collect in
appendix~\ref{sec:typical}.
\par\medskip
\begin{beweis}[of theorem~\ref{thm:ex-and-in}]
  Define the following operators: for
  $j^l\in\fset{T}^l_{\lambda,\delta}$ let
  $$\xi_{j^l}'=\Pi^l_{\rho,\delta}
      \Pi^l_{\hat{\rho},\delta}(j^l)\hat{\rho}_{j^l}\Pi^l_{\hat{\rho},\delta}(j^l)
               \Pi^l_{\rho,\delta}.$$
  We choose $\delta=m\sqrt{\frac{2d}{\epsilon}}$, so that
  \begin{align}
    \label{eq:S:condition}
    S:=\lambda^{\otimes l}(\fset{T}^l_{\lambda,\delta}) &\geq 1-\epsilon, \\
    \label{eq:xi:condition}
    \tr\xi_{j^l}'                                       &\geq 1-\epsilon,
  \end{align}
  which is true by Chebyshev's inequality and
  eqs.~(\ref{eq:cond:typical:prob})
  and~(\ref{eq:typical:prob:general}), specifying $\epsilon$ later.
  \par
  Notice that in this way $\tr\omega'\geq 1-2\epsilon$ for
  $$\omega'=\sum_{j^l\in\fset{T}^l_{\lambda,\delta}} \lambda_{j^l}\xi_{j^l}'.$$
  By eq.~(\ref{eq:typical:rho:lower}) we have
  $$\Pi^l_{\rho,\delta}\omega\Pi^l_{\rho,\delta}\geq \alpha\Pi^l_{\rho,\delta},$$
  with $\alpha=\exp(-lH(\rho)-O(\sqrt{l}))$. Define now $\Pi$ to be the
  projector onto the subspace spanned by the eigenvectors of $\omega'$
  with eigenvalue $\geq \epsilon\alpha$. By construction we find
  $\tr\Omega\geq 1-3\epsilon$ for $\Omega=S^{-1}\Pi\omega'\Pi$.
  \par
  Now let $\xi_{j^l}=\Pi\xi_{j^l}'\Pi$ and define i.i.d.~random variables
  $J^{(\nu)}_\mu\in\fset{T}^l_{\lambda,\delta}$, $\nu=1,\ldots,N$,
  $\mu=1,\ldots,M$ by
  $$\Pr\{J^{(\nu)}_\mu=j^l\}=\frac{\lambda_{j^l}}{S}=:L_{j^l}'.$$
  That is, we consider $N$ independent sets of $M$ independent
  choices each, from $\fset{T}^l_{\lambda,\delta}$.
  Observe that $\Omega=\E\xi_{J^{(\nu)}_\mu}$, the expected
  value of the random operators $\xi_{J^{(\nu)}_\mu}$.
  \par
  We shall show that with high probability the following
  conditions hold:
  \begin{equation}
    \frac{1}{M}\sum_{\mu=1}^M \xi_{J^{(\nu)}_\mu}
                                  \in[(1\pm\epsilon)\Omega],\tag{I${}_\nu$}
  \end{equation}
  for all $\nu$, and
  \begin{equation}
    \frac{1}{NM}\sum_{\nu,\mu=1}^{N,M} \delta_{J^{(\nu)}_\mu}
                                  \in[(1\pm\epsilon)L'].\tag{II}
  \end{equation}
  This is most easily seen with the help of lemma~\ref{lemma:largedev}:
  according to it
  \begin{align*}
    \Pr\{\neg{\rm I}_\nu\}
        &\leq 2\tr\Pi\exp\left(-M\frac{\epsilon^3\alpha}{2\beta\ln 2}\right), \\
    \Pr\{\neg{\rm II}\}
        &\leq 2|\fset{T}^l_{\lambda,\delta}|
                     \exp\left(-NM\frac{\epsilon^2\gamma}{2\ln 2}\right),
  \end{align*}
  with
  $$\gamma=    \min\{\lambda_{j^l}:j^l\in\fset{T}^l_{\lambda,\delta}\}
          \geq \exp\left(-lH(\lambda)-Km\delta\sqrt{l}\right),$$
  compare eq.~(\ref{eq:typical:rho:lower}).
  Choosing $M$ and $N$ according to the theorem's statement
  will force the sum of these probabilities to be less than
  $1$, i.e. with positive probability all the events
  (I${}_\nu$) and (II) happen.
  \par
  Let us assume we fix now values for the $J^{(\nu)}_\mu$ such
  that all equations~(I${}_\nu$) and (II) are satisfied.
  Then we may define operators
  \begin{equation*}\begin{split}
    A^{(\nu)}_{j^l} &=\frac{S}{1+\epsilon}\,\omega^{-1/2}\!
                           \left(
                         \frac{1}{M}\sum_{\mu: J^{(\nu)}_\mu=j^l}\xi_{J^{(\nu)}_\mu}
                           \right) \!\omega^{-1/2}                              \\
                    &=\frac{S}{1+\epsilon}\frac{|\{\mu:J^{(\nu)}_\mu=j^l\}|}{M}
                                             \omega^{-1/2}\xi_{j^l}\omega^{-1/2}.
  \end{split}\end{equation*}
  We check that for each $\nu$ these form a sub--POVM (i.e., a
  collection of positive operators with sum upper bounded by $\1$):
  using (I${}_\nu$) and the definitions of $\Omega$
  and $\omega'$ we find
  \begin{equation*}\begin{split}
    \sqrt{\omega}\left(\sum_{j^l} A^{(\nu)}_{j^l}\right)\sqrt{\omega}
          &=    \frac{S}{1+\epsilon}\frac{1}{M}\sum_{\mu=1}^M \xi_{J^{(\nu)}_\mu} \\
          &\leq S\Omega =    \Pi\omega'\Pi \\
          &\leq \omega' \leq \omega.
  \end{split}\end{equation*}
  Finally, we check that condition (CM) holds:
  it is sufficient to do this for the sub--POVM constructed,
  because then we can distribute the remaining operator weight
  to fill up to $\1$ arbitrarily.
  \par
  We calculate directly from the definitions:
  \begin{equation}\begin{split}
    \sum_{j^l}\frac{1}{2}
           &     \|\sqrt{\omega}(a_{j^l}-A_{j^l})\sqrt{\omega}\|_1                       \\
           &=    \sum_{j^l} \frac{1}{2}\left\|\lambda_{j^l}\hat{\rho}_{j^l}
              -\frac{S|\{\nu\mu:J^{(\mu)}_\mu=j^l\}|}{(1+\epsilon)NM}\xi_{j^l}\right\|_1 \\
           &\leq \frac{1}{2}(1-S)
                +\sum_{j^l\in\fset{T}^l_{\lambda,\delta}}
                    L_{j^l}'\frac{1}{2}\|\hat{\rho}_{j^l}-\xi_{j^l}\|_1                  \\
           &\phantom{=}+\frac{1}{2}\left\|L'
                             -\frac{1}{NM}\sum_{\nu\mu} \delta_{J^{(\nu)}_\mu}\right\|_1 \\
           &\leq \frac{\epsilon}{2}
                +\sum_{j^l\in\fset{T}^l_{\lambda,\delta}}
                    L_{j^l}'\frac{1}{2}\|\hat{\rho}_{j^l}-\xi_{j^l}\|_1
                +\frac{\epsilon}{2}                                                      \\
           &\leq \epsilon+\sum_{j^l\in\fset{T}^l_{\lambda,\delta}}
                    L_{j^l}'\left(\frac{1}{2}\|\hat{\rho}_{j^l}-\xi_{j^l}'\|_1
                                   +\frac{1}{2}\|\xi_{j^l}'-\xi_{j^l}\|_1\right).
    \label{eq:crucial:estimate}
  \end{split}\end{equation}
  By the definition of $\xi_{j^l}'$, using eq.~(\ref{eq:xi:condition})
  and lemma~\ref{lemma:tender:operator} below,
  we can bound the first of the two terms in brackets
  by $\epsilon+\sqrt{2\epsilon}$.
  It remains to estimate the second: consider
  $$\Omega'=\sum_{j^l\in\fset{T}^l_{\lambda,\delta}} L_{j^l}'\xi_{j^l}',$$
  and recall that $\xi_{j^l}=\Pi\xi_{j^l}'\Pi$, hence
  $\Omega=\Pi\Omega'\Pi$.
  By construction we have
  $$\sum_{j^l\in\fset{T}^l_{\lambda,\delta}} L_{j^l}'\tr\xi_{j^l}'
                                                      \geq 1-3\epsilon,$$
  thus, using lemma~\ref{lemma:tender:operator} with each of
  the $\xi_{j^l}'$ and employing concavity of the square root
  function, we end up with
  $$\sum_{j^l\in\fset{T}^l_{\lambda,\delta}} L_{j^l}'
        \frac{1}{2}\|\xi_{j^l}'-\xi_{j^l}\|_1\leq \sqrt{6\epsilon},$$
  which allows us to estimate~(\ref{eq:crucial:estimate})
  by $2\epsilon+\sqrt{2\epsilon}+\sqrt{6\epsilon}$.
\end{beweis}
Here is the lemma that we needed in the proof: it says that a POVM
element that is likely to respond to a state acts ``gently'' on it in the
sense of little disturbance.
\begin{lemma}[Lemma V.9 of~\cite{winter:ieee_strong}]
  \label{lemma:tender:operator}
  For a state $\rho$ and and an operator $0\leq X\leq\1$,
  if $\tr(\rho X)\geq 1-\lambda$, then
  $\left\|\rho-\sqrt{X}\rho\sqrt{X}\right\|_1\leq \sqrt{8\lambda}$.
  \par
  The same is true if $\rho$ is only a subnormalized
  density operator.
  \qed
\end{lemma}

\section{Previous approaches}
\label{sec:upgrade}
The question addressed in the present paper of quantifying
the ``amount of information obtained by a quantum measurement''
has been posed before, in the works~\cite{massar:popescu}
and~\cite{winter:massar:POVMcompr}, with mathematical modellings
different from ours, though there is an evolution leading from
the first to the present:
\par
In~\cite{massar:popescu} the POVM ${\bf a}$ was assumed to maximize a
certain Bayesian gain (there called ``fidelity'')
$$F({\bf a})=\sum_{ij} p_i \tr(\rho_i a_j) F_{ij},$$
to achieve the optimal (i.e.~maximal) value $F_{\rm opt}$.
On blocks of length $l$ the gain (or fidelity) function was extended
by defining $F_{i^lj^l}=\frac{1}{l}\sum_{k=1}^l F_{i_kj_k}$.
This definition has the easily checked property that the gain
on blocks of length $l$,
\begin{equation}
  \label{eq:block:fidelity}
  F({\bf a}^{\otimes l})=
      \sum_{i^lj^l} p_{i^l} \tr(\rho_{i^l}a_{j^l}) F_{i^lj^l},
\end{equation}
equals the single letter expression $F({\bf a})$.
\par
Note that in this way the maximum Bayesian gain is still $F_{\rm opt}$
(which can be seen from eq.~(\ref{eq:POVM:fidelity}) below).
Then the following theorem was shown:
\begin{thm}[Massar,~Popescu~\cite{massar:popescu}]
  \label{thm:massar:popescu}
  For $\epsilon>0$ and $l$ large enough there exists a POVM ${\bf A}$
  with fidelity $F({\bf A})\geq F_{\text{opt}}-\epsilon$ and
  $$M\leq\exp(l(H(\rho)+\epsilon))$$
  many outcomes among the $j^l$.
  \qed
\end{thm}
This result was interpreted as saying that about any property of the ensemble states,
as encoded in the Bayesian gain matrix $F_{ij}$, one can learn at
most one bit per qubit.
\par
In~\cite{winter:massar:POVMcompr} this was extended and clarified as follows:
observe that for any POVM ${\bf A}=(A_{j^l_\mu})_{\mu=1,\ldots,M}$ one has
\begin{equation}\begin{split}
  \label{eq:POVM:fidelity}
  F({\bf A}) &=\sum_{i^l} p_{i^l}\sum_\mu \tr(\rho_{i^l}A_{j^l_\mu})
                                           \frac{1}{l}\sum_{k=1}^l F_{i_kj_{\mu k}} \\
             &=\frac{1}{l}\sum_{k=1}^l\sum_i\sum_j p_i \tr(\rho_i ({\bf A}|k)_j) F_{ij},
\end{split}\end{equation}
where (with $[l]=\{1,\ldots,l\}$)
\begin{equation}\begin{split}
  \label{eq:POVM:marginals}
  ({\bf A}|k)_j &=\tr_{\neq k}\left(\left(\rho^{\otimes [l]\setminus k}\otimes\1_k\right)
                         \sum_{\mu:\ j_{\mu k}=j}A_{j^l_\mu}\right)   \\
          &=\rho^{-1}\tr_{\neq k}\left(\rho^{\otimes l}
                         \sum_{\mu:\ j_{\mu k}=j}A_{j^l_\mu}\right)   \\
          &\hspace{-.5cm}=\sqrt{\rho^{-1}}\tr_{\neq k}\!\left(\sqrt{\rho}^{\otimes l}
                  \left(\sum_{\mu:\ j_{\mu k}=j}A_{j^l_\mu}\right)
                                    \sqrt{\rho}^{\otimes l}\right)\!\sqrt{\rho^{-1}}.
\end{split}\end{equation}
For each $k$, the collection $(({\bf A}|k)_j)_{j=1,\ldots,m}$
obviously is a POVM on ${\cal H}$. We may assume (as we
shall do in the sequel) that the $|F_{ij}|$
are bounded by $1$: then the fidelity condition of
theorem~\ref{thm:massar:popescu}, reading
\begin{equation}
  |F({\bf A})-F({\bf a})|\leq\epsilon,\tag{C0}
\end{equation}
is implied by
\begin{equation}
  \forall k\quad \sum_{ij}
        \left| p_i\tr\left(\rho_i ({\bf A}|k)_j\right)-p_i\tr(\rho_i a_j)\right|
                                     \leq\epsilon.\tag{C1}
\end{equation}
This is itself implied by
\begin{equation}
  \forall k\forall i\ \sum_j |\tr(\rho_i ({\bf A}|k)_j)-\tr(\rho_i a_j)|
                                      \leq\epsilon,\tag{C2}
\end{equation}
which in turn follows from
\begin{equation}
  \forall k\ \sum_j \|({\bf A}|k)_j-a_j\|\leq\epsilon.\tag{C3}
\end{equation}
It was then proved
\begin{thm}[Winter,~Massar~\cite{winter:massar:POVMcompr}]
  \label{thm:POVM:compr:1}
  For the state $\rho$ and the POVM ${\bf a}$ define a canonical
  ensemble $\{\hat{\rho}_j,\lambda_j\}$, with states
  $$\hat{\rho}_j=\frac{1}{\tr(\rho a_j)}{\sqrt{\rho}a_j\sqrt{\rho}}$$
  and probabilities $\lambda_j=\tr(\rho a_j)$.
  \par
  Given $\epsilon>0$, there exists a POVM
  ${\bf A}=(A_{j^l_\mu})_{\mu=1,\ldots,M}$ with
  $$M\leq
    \exp\left(l\left(H(\rho)-\sum_j \lambda_j H(\hat{\rho}_j)\right)+C\sqrt{l}\right)$$
  (where $C$ is a constant depending only on $\epsilon$, $d$ and $m$), and such that
  (C3) is satisfied.
  \qed
\end{thm}
This theorem is in an asymptotic sense best possible (such an
optimality was missing in~\cite{massar:popescu}):
\begin{thm}[Winter,~Massar~\cite{winter:massar:POVMcompr}]
  \label{thm:POVM:compr:lower}
  Let $0<\epsilon\leq (\lambda_0/2)^2$, with
  $\lambda_0=\min_j \lambda_j$. Then for any POVM
  ${\bf A}=(A_{j^l_\mu})_{\mu=1,\ldots,M}$ such that
  (C3) holds, one has
  $$M\geq \exp\left(l\left(H(\rho)-\sum_j \lambda_j H(\hat{\rho}_j)
     +\frac{3\epsilon}{\lambda_0^2}\log\frac{2\epsilon}{\lambda_0^2 d}\right)\right).$$
  \qed
\end{thm}
\begin{figure}[ht]
  \includegraphics{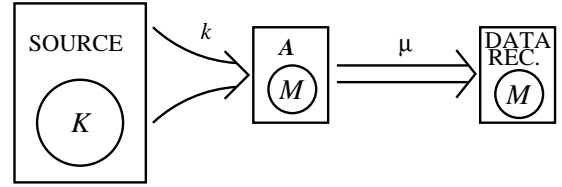}
  \caption{In~\cite{massar:popescu} and~\cite{winter:massar:POVMcompr}
           the original POVM is replaced by an ``equivalent'' one
           (as made precise in theorems~\ref{thm:massar:popescu}
           and~\ref{thm:POVM:compr:1}) with much fewer
           outcomes. So, POVM and data record need much less rate
           of processing and storage, respectively. Of course,
           compared to theorem~\ref{thm:ex-and-in} we loose many
           potential measurement results in constructing the new POVM.}
  \label{fig:compress}
\end{figure}
\par
Here we want to show that the theorems~\ref{thm:massar:popescu}
and~\ref{thm:POVM:compr:1} may be obtained as corollaries of 
theorem~\ref{thm:ex-and-in}.
\par
\begin{beweis}[of theorem~\ref{thm:massar:popescu}]
  Choose $x_\nu$ and ${\bf A}^{(\nu)}$ according to
  theorem~\ref{thm:ex-and-in}, such that condition (CM)
  is satisfied for ${\bf A}=\sum_\nu x_\nu {\bf A}^{(\nu)}$,
  with some $\epsilon>0$ (which implies
  that also (CP) is satisfied with the same $\epsilon$).
  Then, assuming without loss of generality that
  $|F_{ij}|\leq 1$, we get immediately out of
  eq.~(\ref{eq:block:fidelity}) that
  $$|F({\bf A})-F({\bf a}^{\otimes l})|\leq \epsilon.$$
  Since we assume that ${\bf a}$ maximizes $F$ we conclude,
  using linearity of $F$ in the POVM:
  \begin{equation*}\begin{split}
    F_{\rm opt}-\epsilon &=    F({\bf a}^{\otimes l})-\epsilon \\
                         &\leq F({\bf A}) = \sum_\nu x_\nu F({\bf A}^{(\nu)}).
  \end{split}\end{equation*}
  This finally means that for at least one $\nu$
  $$F({\bf A}^{(\nu)})\geq F_{\rm opt}-\epsilon,$$
  which is what we wanted to prove: recall that ${\bf A}^{(\nu)}$
  has $M\leq \exp(lI(\lambda;\hat{\rho})+O(\sqrt{l}))
         \leq \exp(lH(\rho)+O(\sqrt{l}))$ many outcomes.
\end{beweis}
Note that the latter estimate is met
with equality if and only if ${\bf a}$ is maximally refined
(i.e., consists of rank--$1$ operators only), so regardless
of ${\bf a}$, $H(\rho)$ is \emph{the} rate of intrinsic data
of any probing of the ensemble states.
\par
Note further that our derivation does not depend on the
particular structure of the block--fidelity: obviously
we can as well conclude for any ensemble
$\{\sigma_k,q_k\}$ with average $\omega$ and any
fidelity matrix $F_{k j^l}$ that
\begin{equation*}\begin{split}
  \left|F({\bf A})-F({\bf a}^{\otimes l})\right|
     &\leq \sum_{k j^l} q_k |\tr(\sigma_k A_{j^l})-\tr(\sigma_k a_{j^l})| F_{k j^l} \\
     &\leq \epsilon \|F\|,
\end{split}\end{equation*}
with $\|F\|:=\max_{k j^l} |F_{k j^l}|$. If now
$\|F\|\leq O(F({\bf a}^{\otimes l}))$ for $l\rightarrow\infty$
then we get (for sufficiently large $l$)
$$F({\bf A})\geq (1-\epsilon)F({\bf a}^{\otimes l}).$$
\par\medskip
Of course, as explained in the introduction,
theorem~\ref{thm:massar:popescu} is really a corollary of
theorem~\ref{thm:POVM:compr:1}. So, we continue to prove
the latter:
\par
\begin{beweis}[of theorem~\ref{thm:POVM:compr:1}]
  Assume that a collection of POVMs ${\bf A}^{(\nu)}$,
  $\nu=1,\ldots,N$ like in theorem~\ref{thm:ex-and-in}
  is chosen, with probabilities $x_\nu$, such that
  ${\bf A}'=\sum_\nu x_\nu {\bf A}^{(\nu)}$ satisfies (CM).
  Define i.i.d.~random variables $T_1,\ldots,T_Q$,
  each with $\Pr\{T_q=\nu\}=x_\nu$.
  We want to study the random POVMs ${\bf A}^{(T_q)}$,
  and especially their mean
  $${\bf A}=\frac{1}{Q}\sum_{q=1}^Q {\bf A}^{(T_q)}.$$
  Observe that $\E{\bf A}=\E{\bf A}^{(T_q)}={\bf A}'$.
  \par
  Recall the definition of marginal POVMs. Obviously,
  by linearity of this definition, we have
  $$({\bf A}|k)=\frac{1}{Q}\sum_{q=1}^Q ({\bf A}^{(T_q)}|k)$$
  and
  $$\E({\bf A}|k)=\E({\bf A}^{(T_q)}|k)=({\bf A}'|k).$$
  From condition (CM) and the monotonicity of the trace
  norm under partial trace we get now, for every $k$,
  \begin{equation}
    \label{eq:part1}
    \sum_j \frac{1}{2}\left\|\sqrt{\rho}(({\bf A}'|k)_j-a_j)\sqrt{\rho}\right\|_1
                    \leq\epsilon.
  \end{equation}
  Denoting the smallest nonzero
  eigenvalue of any of the $\sqrt{\rho}a_j\sqrt{\rho}$
  by $u$, and choosing $\epsilon$ small enough, this assures
  that $\sqrt{\rho}({\bf A}'|k)_j\sqrt{\rho}$
  \emph{restricted to the support of} $\sqrt{\rho}a_j\sqrt{\rho}$
  is lower bounded by $u/2$.
  Then we can apply lemma~\ref{lemma:largedev} and obtain
  \begin{equation*}\begin{split}
    \Pr &\Bigl\{\sqrt{\rho}({\bf A}|k)_j\sqrt{\rho} 
           \not\in \bigl[(1\pm\epsilon)\sqrt{\rho}({\bf A}'|k)_j\sqrt{\rho}\bigr]
                   \text{ on }\supp\hat{\rho}_j\Bigr\}                            \\
        &\hspace{4.7cm}  \leq 2d \exp\left(\!-Q\frac{\epsilon^2 u}{4\ln 2}\right)\!.
  \end{split}\end{equation*}
  Thus we can estimate the sum of these probabilities over all
  $k=1,\ldots,l$ and $j=1,\ldots,m$ to less than $1$ if
  $$Q\geq 1+\frac{4\ln 2}{\epsilon^2 u}\log(2dlm).$$
  This implies that there exist actual values of the $T_q$ such that
  for all $k$
  \begin{equation}
    \label{eq:local:1}
    \sum_j \frac{1}{2}\left\|\sqrt{\rho}({\bf A}|k)_j\sqrt{\rho}|_{\supp\hat{\rho}_j}
                                                     -\sqrt{\rho}a_j\sqrt{\rho}\right\|_1
                                                                              \leq 2\epsilon,
  \end{equation}
  where we observed that the $\sqrt{\rho}({\bf A}'|k)_j\sqrt{\rho}$
  all have trace at most $1$, and have used eq.~(\ref{eq:part1}).
  Hence we get (with $\Lambda_{kj}=\tr(\rho({\bf A}|k)_j)$ and
  $\Lambda_{kj}\widehat{\Rho}_{kj}=\sqrt{\rho}({\bf A}|k)_j\sqrt{\rho}$)
  $$\sum_j \Lambda_{kj}\tr\left(\widehat{\Rho}_{kj}|_{\supp\hat{\rho}_j}\right) \geq 1-2\epsilon,$$
  and using lemma~\ref{lemma:tender:operator}, this gives
  \begin{equation}
    \label{eq:local:2}
    \sum_j \frac{1}{2}\left\|\sqrt{\rho}({\bf A}|k)_j\sqrt{\rho}|_{\supp\hat{\rho}_j}
                              -\sqrt{\rho}({\bf A}|k)_j\sqrt{\rho}\right\|_1\leq 2\sqrt{\epsilon}.
  \end{equation}
  Now~(\ref{eq:local:1}) and~(\ref{eq:local:2}) yield
  $$\sum_j \frac{1}{2}\left\|\sqrt{\rho}(({\bf A}|k)_j-a_j)\sqrt{\rho}\right\|_1
                                                      \leq 2\epsilon+2\sqrt{\epsilon},$$
  Denoting the minimal eigenvalue of $\rho$ by $r$ (which we assumed
  to be positive) this readily implies
  $$\sum_j \frac{1}{2}\left\|({\bf A}|k)_j-a_j\right\|\leq
                                                \frac{2\epsilon+2\sqrt{\epsilon}}{rd},$$
  and we are done, since ${\bf A}$ has only $MQ$ many possible outcomes.
\end{beweis}

\section{Strong converse}
\label{sec:strong:converse}
In this section we prove the asymptotic optimality of the separation of the
measurement from theorem~\ref{thm:ex-and-in}. To be precise, it is
\begin{thm}
  \label{thm:strong:converse}
  Whenever there are POVMs ${\bf A}^{(\nu)}$ on $[m]^l$, $\nu=1,\ldots,N$,
  each supported on at most $M$ elements, and probability weights $x_\nu>0$, such that
  ${\bf A}=\sum_\nu x_\nu {\bf A}^{(\nu)}$ satisfies condition (CM),
  for some $\epsilon<1$, then
  \begin{align*}
    M  &\geq \exp \left( l I(\lambda;\hat{\rho})-O(\sqrt{l}) \right),\\
    MN &\geq \exp \left( l H(\lambda)-O(\sqrt{l}) \right),
  \end{align*}
  where the constants depend only on $\epsilon$.
\end{thm}
\begin{beweis}
  Let us begin with the second inequality:
  by construction the set $\fset{R}\subset[m]^l$ of possible
  outcomes of ${\bf A}$ has cardinality at most $MN$.
  Denoting by $\Lambda$ the distribution of outcomes
  according to ${\bf A}$, i.e.
  $$\Lambda_{j^l}=\tr\left( \rho^{\otimes l}A_{j^l}\right),$$
  from (CM) we get immediately
  \begin{equation}
    \label{eq:distributions:close}
    \frac{1}{2}\|\lambda^{\otimes l}-\Lambda\|_1\leq\epsilon,
  \end{equation}
  which in turn implies
  $$\lambda^{\otimes l}(\fset{R})\geq \Lambda(\fset{R})-\epsilon
                                 =    1-\epsilon.$$
  By a well known trick~\cite{wolfowitz:coding}
  the lower bound now follows: we consider
  $\fset{R}'=\fset{R}\cap\fset{T}^l_{\lambda,\delta}$,
  with $\delta=\sqrt{\frac{1-\epsilon}{2m}}$, whence
  we have, using Chebyshev's inequality
  $$\lambda^{\otimes l}(\fset{R}')\geq \frac{1-\epsilon}{2}.$$
  Using the fact (compare eq.~(\ref{eq:typical:rho:lower}))
  $$\forall j^l\in\fset{T}^l_{\lambda,\delta}\quad
      \lambda_{j^l} \leq \exp\left(-lH(\lambda)+Km\delta\sqrt{l}\right),$$
  we conclude
  $$MN\geq|\fset{R}|\geq|\fset{R}'|
      \geq\frac{1-\epsilon}{2}\exp\left(lH(\lambda)-Km\delta\sqrt{l}\right).$$
  \par
  Now for the first inequality: introduce the ensembles
  $\{\widehat{\Rho}^{(\nu)}_{j^l},\Lambda^{(\nu)}_{j^l}\}_{j^l}$ with
  $$\Lambda^{(\nu)}_{j^l}\widehat{\Rho}^{(\nu)}_{j^l}
                        =\sqrt{\omega}A^{(\nu)}_{j^l}\sqrt{\omega},$$
  all of which have average $\omega$. Then we define the (subnormalized)
  density operators
  \begin{align*}
    \tilde{\rho}_{j^l}
      &=\Pi^l_{\hat{\rho},\delta}(j^l)\rho_{j^l}\Pi^l_{\hat{\rho},\delta}(j^l), \\
    \widetilde{\Rho}^{(\nu)}_{j^l}
      &=\Pi^l_{\hat{\rho},\delta}(j^l)\Rho^{(\nu)}_{j^l}\Pi^l_{\hat{\rho},\delta}(j^l),
  \end{align*}
  for $j^l\in\fset{T}^l_{\lambda,\delta}$, with
  $\delta=\sqrt{\frac{4mn}{1-\epsilon}}$.
  Then by Chebyshev inequality and eq.~(\ref{eq:cond:typical:prob})
  $$\frac{1}{2}\left\|\omega-
           \sum_{j^l\in\fset{T}^l_{\lambda,\delta}}\!
                   \lambda_{j^l}\tilde{\rho}_{j^l}\right\|_1
                                           \leq \frac{1-\epsilon}{2},$$
  while from (CM) we get
  \begin{equation}
    \label{eq:states:close}
    \frac{1}{2}\left\|\sum_{j^l\in\fset{T}^l_{\lambda,\delta}}\!\!
                           \lambda_{j^l}\tilde{\rho}_{j^l}
                      -\sum_\nu x_\nu \!\! \sum_{j^l\in\fset{T}^l_{\lambda,\delta}}\!\!
                        \Lambda^{(\nu)}_{j^l}\widetilde{\Rho}^{(\nu)}_{j^l}\right\|_1
                                                   \leq \frac{1+\epsilon}{2}=:\epsilon'.
  \end{equation}
  These immediately imply
  \begin{equation}
    \label{eq:average:trace}
    \sum_\nu x_\nu \! \sum_{j^l\in\fset{T}^l_{\lambda,\delta}}
                          \Lambda^{(\nu)}_{j^l}\tr\widetilde{\Rho}^{(\nu)}_{j^l}
                                                  \geq 1-\epsilon',
  \end{equation}
  so there exists at least one $\nu$ such that
  $$\sum_{j^l\in\fset{T}^l_{\lambda,\delta}}
        \Lambda^{(\nu)}_{j^l}\tr\widetilde{\Rho}^{(\nu)}_{j^l} \geq 1-\epsilon'.$$
  Now consider the (subnormalized) density operators
  $$\Theta^{(\nu)}_{j^l}=\sqrt{\widehat{\Rho}^{(\nu)}_{j^l}}
                                  \Pi^l_{\hat{\rho},\delta}
                         \sqrt{\widehat{\Rho}^{(\nu)}_{j^l}},$$
  which evidently satisfy
  $$\theta:=   \sum_{j^l\in\fset{T}^l_{\lambda,\delta}}
                         \Lambda^{(\nu)}_{j^l}\Theta^{(\nu)}_{j^l}
          \leq \sum_{j^l} \Lambda^{(\nu)}_{j^l}\widehat{\Rho}^{(\nu)}_{j^l}
          =    \omega.$$
  Denoting with $\Pi$ the projection onto the support of $\theta$
  and inserting $\tr\Theta^{(\nu)}_{j^l}=\tr\widetilde{\Rho}^{(\nu)}_{j^l}$,
  we arrive at
  $$\tr(\omega\Pi)\geq 1-\epsilon',$$
  from where we conclude
  $$\rank\Pi=\tr\Pi\geq \exp\left(lH(\rho)-O(\sqrt{l})\right).$$
  This follows by a standard reasoning (which we take
  from~\cite{winter:ieee_strong}):
  for $F=\Pi^l_{\rho,\delta}\Pi\Pi^l_{\rho,\delta}$, choosing
  $\delta$ large enough, we get
  $$\tr\left(\Pi^l_{\rho,\delta}\omega\Pi^l_{\rho,\delta}\,\Pi\right)
                                  =\tr(\omega F)\geq \frac{1-\epsilon'}{2}.$$
  By eq.~(\ref{eq:typical:rho:lower}) the inequality follows.
  \par
  On the other hand each of the $\Theta^{(\nu)}_{j^l}$ has
  rank at most $\exp\left(lH(\hat{\rho}|P)+O(\sqrt{l})\right)$,
  and we deduce our claim.
\end{beweis}
\par
We may relax a bit the condition of the theorem regarding
the parameter $M$:
if we allow the different POVMs ${\bf A}^{(\nu)}$ to
have different numbers $M_\nu$ of possible outcomes,
then we can prove the slightly stronger estimate
$$M:=\sum_\nu x_\nu M_\nu\geq \exp\left( l I(\lambda;\hat{\rho})-O(\sqrt{l}) \right).$$
(while the second inequality obviously holds for $\sum_\nu M_\nu$).
To see this go back to eq.~(\ref{eq:average:trace}) and observe that
by a Markov inequality argument
$$\Pr\nolimits_x\left\{\nu:\sum \Lambda^{(\nu)}_{j^l}\tr\widetilde{\Rho}^{(\nu)}_{j^l}
                                                  \geq 1-\sqrt{\epsilon'}\right\}
                                                           \geq 1-\sqrt{\epsilon'},$$
whence the claim directly follows.
\begin{rem}
  \label{rem:CP}
  While in the above proof we assumed the property (CM) for
  $\epsilon<1$, we conjecture that (CP) for all sources with average $\omega$,
  with $\epsilon<1$, is sufficient to arrive at its conclusion.
  \par
  Let us inspect this possibility along the lines of the proof:
  crucial were the estimates~(\ref{eq:distributions:close})
  and~(\ref{eq:states:close}), the former being an immediate consequence
  of (CP), so we would have to show this only for the latter.
  However, this demonstration has escaped us so far.
  \par\medskip
  Finally, a comment on why this converse is \emph{strong}: optimality
  of theorem~\ref{thm:ex-and-in} is proved already by our observation
  in the previous section that it implies theorem~\ref{thm:POVM:compr:1},
  and the lower bound of theorem~\ref{thm:POVM:compr:lower}.
  However, closer inspection of this lower bound reveals that
  it coincides with the upper bound only in the limit $\epsilon\rightarrow 0$.
  For positive $\epsilon$ it leaves room for a tradeoff between compression
  and error (not untypical for the type of error concept we had used).
  This is known in information theory as a
  \emph{weak converse}~\cite{wolfowitz:coding}.
  The strong converse in contrast shows optimality of the upper bound
  in the asymptotic limit $l\rightarrow\infty$, with any $\epsilon$
  bounded away from $1$.
\end{rem}

\section{Asymptotic decomposition of\protect\\ instruments and operations}
\label{sec:other:appl}
An interesting generalization of our main theorem
arises from the point of view that POVMs
are just a special case of general open dynamics: the most general
form of evolution is a completely positive, trace
preserving linear map
$\varphi_*$ from states on ${\cal H}$ to states on ${\cal K}$.
Such a map can (non--uniquely) be represented in the Kraus form
\begin{equation}
  \label{eq:kraus}
  \varphi_*:\pi\longmapsto \sum_{j=1}^m V_j\pi V_j^*,
\end{equation}
where $V_j:{\cal H}\rightarrow{\cal K}$ are $\C$--linear
and $\sum_j V_j^* V_j=\1$. The representation can be made unique
by considering it as a partial measurement, and including the outcome
$j$: extend the output system to ${\cal K}\otimes{\cal J}$,
and modify the map $\varphi_*$ to
$$\widetilde{\varphi}_*:\pi\longmapsto \sum_{j=1}^m V_j\pi V_j^* \otimes\ketbra{j}.$$
(Technically this will amount to a change of the Kraus operators, too,
but we will not need the details here). This is the notion
of an \emph{instrument} (Davies and Lewis~\cite{davies:lewis:instrument}).
One can see that it is representable in Kraus form, too, so
we will in the sequel always look at a particular Kraus representation.
\par
In analogy to the question about POVMs of this
work we would like to approximate $\varphi_*^{\otimes l}$
by the average of some $\Phi^{(\nu)}_*$, $\nu=1,\ldots,N$,
each of which should have a Kraus representation
with a small number of contributing operators. As is well known
this number is the dimension of the ancillary system
(environment) sufficient
to emulate the effect of the operation by a unitary interaction
and subsequent partial trace. Its logarithm is an upper bound
on the ``information leakage'' from the system
to the environment.
\par
Note that (apart from looking at approximation)
we are considering here the
problem of convex decomposition of completely positive maps,
like we did before for POVMs.
Of course, every completely positive map
has a decomposition into extremal such ones, with possibly
fewer terms in the Kraus representation. For this one can employ
a theorem of Choi~\cite{choi:cp}, saying that $\varphi_*$
from eq.~(\ref{eq:kraus}) is extremal if and only if
the family of operators $V_j^* V_k$ is linearly independent
(in particular, then $m\leq d$).
\par
We show now how to solve this problem as a consequence of
theorem~\ref{thm:ex-and-in}, with an additional reasoning mainly
directed to quantum state fidelities:
\par
Formally, we are looking for a family of maps
\begin{equation}\begin{split}
  \label{eq:big:cp-map}
  \Phi^{(\nu)}_*:{\cal B}({\cal H}^{\otimes l})
                         &\longrightarrow {\cal B}({\cal K}^{\otimes l}),\\
         \sigma          &\longmapsto \sum_{\mu=1}^M W^{(\nu)}_\mu \sigma W^{(\nu)*}_\mu
\end{split}\end{equation}
and probabilities $x_\nu$ such that for
$\Phi_*=\sum_\nu x_\nu \Phi^{(\nu)}_*$ and any ensemble
$\{\sigma_k,q_k\}$ with average $\omega=\rho^{\otimes l}$
the following condition holds:
\begin{equation}
  \sum_k q_k \frac{1}{2}\left\|\varphi^{\otimes l}_*(\sigma_k)
                                    -\Phi_*(\sigma_k)\right\|_1\leq\epsilon.\tag{CO}
\end{equation}
In fact, there is an appealing way to state them all together,
and strengthen the content at the same time:
for a purification $\pi$ of $\rho$ on an extended system
${\cal H}\otimes{\cal H}'$ we ask for
\begin{equation}
  \frac{1}{2}\left\|(\varphi_*\otimes\id)^{\otimes l}(\pi^{\otimes l})
                    -(\Phi_*\otimes\id^{\otimes l})(\pi^{\otimes l})\right\|_1\leq \epsilon.\tag{CO*}
\end{equation}
Indeed, this implies (CO): just observe that by choosing a POVM $(T_k)$
on ${\cal H}^{\prime \otimes l}$ one can ``induce'' any ensemble $\{\sigma_k,q_k\}$
on ${\cal H}^{\otimes l}$ for $\omega$, in the following sense:
$$q_k\sigma_k=\tr_{{\cal H}^{\prime \otimes l}} \left( \pi^{\otimes l}(\1\otimes T_k)\right).$$
How to do this is explained in detail in~\cite{hjw:classification}
(or see appendix A below).
Note that this generalizes the implication of (CP) from (CM), discussed
earlier, when we view the POVMs as the quantum operations
eqs.~(\ref{eq:POVM:op1}) and~(\ref{eq:POVM:op2}).
\par
Conversely, assuming (CO) for all ensembles for $\omega$ does unfortunately
not imply (CO*) with a comparable error parameter. (Examples are not hard to
construct for which (CO) holds with a small $\epsilon$ while the bound
in (CO*) is close to $1$.)
\par\medskip
With $\varphi_*$ there is associated the POVM
$${\bf a}=(a_j=V_j^*V_j:i=j,\ldots,m),$$
and with this goes the ensemble $\{\hat{\rho}_j,\lambda_j\}$, as before.
\begin{thm}
  \label{thm:cp-map}
  With the above notation and $\epsilon>0$ there exist
  quantum operations in the form of eq.~(\ref{eq:big:cp-map}), with
  \begin{align*}
    M &\leq \exp\left(l I(\lambda;\hat{\rho})+O(\sqrt{l})\right), \\
    N &\leq \exp\left(l (H(\lambda)-I(\lambda;\hat{\rho}))+O(\sqrt{l})\right),
  \end{align*}
  and such that $\Phi_*=\frac{1}{N}\sum_\nu \Phi^{(\nu)}_*$
  satisfies (CO*).
  \par
  These bounds are asymptotically best possible if $\varphi_*$
  is an instrument.
\end{thm}
\begin{beweis}
  Let ${\bf A}^{(\nu)}$ and $x_\nu$ be the POVMs and probabilities
  constructed in theorem~\ref{thm:ex-and-in} from ${\bf a}^{\otimes l}$
  and $\omega=\rho^{\otimes l}$, and let ${\bf A}=\sum_\nu x_\nu {\bf A}^{(\nu)}$.
  We use the notation from the proof of this theorem and from
  section~\ref{sec:strong:converse}:
  \begin{align*}
    \sqrt{\rho}a_j\sqrt{\rho} &=\lambda_j\hat{\rho}_j, \\
    \sqrt{\omega}A^{(\nu)}_{j^l}\sqrt{\omega} 
                              &=\Lambda^{(\nu)}_{j^l}\widehat{\Rho}^{(\nu)}_{j^l}, \\
    \Lambda_{j^l}             &=\sum_\nu x_\nu \Lambda^{(\nu)}_{j^l}.
  \end{align*}
  Note that by the proof of theorem~\ref{thm:ex-and-in}
  the $\widehat{\Rho}^{(\nu)}_{j^l}$ either are $0$ or
  equal to $\widehat{\Rho}_{j^l}:=\frac{S}{1+\epsilon}\xi_{j^l}$.
  \par
  Introduce the unitaries $U_j$ by the polar decomposition
  \begin{equation}
    \label{eq:polar:U}
    V_j\sqrt{\rho}=U_j\sqrt{\sqrt{\rho}V_j^*V_j\sqrt{\rho}}
                  =U_j\sqrt{\lambda_j\hat{\rho}_j},
  \end{equation}
  and let $U_{j^l}=U_{j_1}\otimes\cdots\otimes U_{j_l}$.
  Now define $W^{(\nu)}_{j^l}$ by letting
  \begin{equation}
    \label{eq:define:W}
    W^{(\nu)}_{j^l}\sqrt{\omega}
           =U_{j^l}\sqrt{\Lambda^{(\nu)}_{j^l}\widehat{\Rho}^{(\nu)}_{j^l}},
  \end{equation}
  and observe that for fixed $\nu$ only $M$ of them are nonzero,
  and that for fixed $j^l$ these are all multiples of each other.
  Hence these operators define a quantum operation $\Phi^{(\nu)}_*$
  according to the theorem, and $\Phi_*=\sum_\nu x_\nu\Phi^{(\nu)}_*$.
  \par
  With these definitions we check that (CO*) is satisfied: using
  $\pi^{\otimes l}=(\sqrt{\omega}\otimes\1^{\otimes l})\ketbra{I}(\sqrt{\omega}\otimes\1^{\otimes l})$
  (see lemma~\ref{lemma:canonical:purifications} in the appendix) and
  eqs.~(\ref{eq:polar:U}) and~(\ref{eq:define:W}) we calculate
  \begin{equation}\begin{split}
    &\left\|(\varphi_*^{\otimes l}\otimes\id^{\otimes l})(\pi^{\otimes l})
                    -(\Phi_*\otimes\id^{\otimes l})(\pi^{\otimes l}) \right\|_1  \\
    &\phantom{==}
     \leq \sum_{j^l} \left\|\left(V_{j^l}^{\phantom{*}}\otimes\1^{\otimes l}\right)\!
                                 \pi^{\otimes l}\!\left(V^*_{j^l}\otimes\1^{\otimes l}\right)\right. \\
    &\phantom{======}
                            -\sum_\nu x_\nu \left(W^{(\nu)}_{j^l}\otimes\1^{\otimes l}\right)\!
                                 \pi^{\otimes l}\!\left(W^{(\nu)*}_{j^l}\otimes\1^{\otimes l}\right)
                     \Bigr\|_1                                                                       \\
    &\phantom{==}
     =    \sum_{j^l} \biggl\|\lambda_{j^l}\left(\sqrt{\hat{\rho}_{j^l}}\otimes\1^{\otimes l}\right)\!
                                 \ketbra{I}\!\left(\sqrt{\hat{\rho}_{j^l}}\otimes\1^{\otimes l}\right) \\
    &\phantom{======}
                            -\Lambda_{j^l}\left(\sqrt{\widehat{\Rho}_{j^l}}\otimes\1^{\otimes l}\right)\!
                                 \ketbra{I}\!\left(\sqrt{\widehat{\Rho}_{j^l}}\otimes\1^{\otimes l}\right)
                     \biggr\|_1                                                                      \\
    &\phantom{==}
     \leq \|\lambda^{\otimes l}-\Lambda\|_1                                                          \\
    &\phantom{===}
     +    \sum_{j^l} \lambda_{j^l}\biggl\|\left(\sqrt{\hat{\rho}_{j^l}}\otimes\1^{\otimes l}\right)\!
                                 \ketbra{I}\!\left(\sqrt{\hat{\rho}_{j^l}}\otimes\1^{\otimes l}\right) \\
    &\phantom{=========}
                          -\!\left(\sqrt{\widehat{\Rho}_{j^l}}\otimes\1^{\otimes l}\right)\!
                                 \ketbra{I}\!\left(\sqrt{\widehat{\Rho}_{j^l}}\otimes\1^{\otimes l}\right)
                     \biggr\|_1 \!.
    \label{eq:long:estimate}
  \end{split}\end{equation}
  The last line here is estimated as follows:
  the first term is bounded by $2\epsilon$
  (see the proof of theorem~\ref{thm:strong:converse}),
  and for the other we use lemma~\ref{lemma:canonical:purifications}:
  observe that for each $j^l$
  the two terms inside the trace norm are the canonical purifications
  of $\hat{\rho}_{j^l}$ and $\widehat{\Rho}_{j^l}$, respectively.
  Thus we get
  \begin{equation*}\begin{split}
    &\biggl\|\left(\sqrt{\hat{\rho}_{j^l}}\otimes\1^{\otimes l}\right)\!
                                 \ketbra{I}\!\left(\sqrt{\hat{\rho}_{j^l}}\otimes\1^{\otimes l}\right) \\
    &\phantom{======;}
                            -\left(\sqrt{\widehat{\Rho}_{j^l}}\otimes\1^{\otimes l}\right)\!
                                 \ketbra{I}\!\left(\sqrt{\widehat{\Rho}_{j^l}}\otimes\1^{\otimes l}\right)
                     \biggr\|_1                                                                      \\
    &\phantom{==================;}
     \leq 2\sqrt{2}\sqrt[4]{\left\|\hat{\rho}_{j^l}-\widehat{\Rho}_{j^l}\right\|_1},
  \end{split}\end{equation*}
  and using concavity of the root function and the estimate of
  eq.~(\ref{eq:crucial:estimate}) we can upper bound the last line
  of eq.~(\ref{eq:long:estimate}) by $O(\epsilon^{1/8})$.
  \par
  If $\varphi_*$ is an instrument any approximate convex decomposition
  of $\varphi_*^{\otimes l}$ implies a similar decomposition for
  the POVM ${\bf a}^{\otimes l}$. Hence theorem~\ref{thm:strong:converse}
  gives the optimality of the bounds for $M$ and $N$.
\end{beweis}
\par
Interestingly, the bounds of theorem~\ref{thm:cp-map} depend on the
Kraus representation~(\ref{eq:kraus}) of the map $\varphi_*$:
all other such representations are related by unitary transforms,
i.e.
$$\varphi_*(\sigma)=\sum_J V_J'\sigma V_J^{\prime *}$$
if and only if
$$V_J'=\sum_j {\cal U}_{Jj}V_j,$$
with a unitary matrix $({\cal U}_{Jj})_{Jj}$ of
complex numbers. (This is essentially
a consequence of the uniqueness up to unitaries
of the Stinespring dilation~\cite{stinespring}
of $\varphi$, which implies the Kraus representation.
This fact is also discussed in detail in~\cite{schumacher:exchange}).
\par
This motivates the introduction of
\begin{equation}
  \label{eq:sigma}
  \Sigma(\rho;\varphi_*):=\min_{\text{Kraus repr. of }\varphi_*} I(\lambda;\hat{\rho}),
\end{equation}
i.e. the minimum rate of the parameter $M$ in decompositions of
$\varphi_*$ according to theorem~\ref{thm:cp-map}.
\par
Note that, according to~\cite{schumacher:exchange},
the minimum of $H(\lambda)$ over all Kraus representations
is exactly $S_e$, the \emph{entropy exchange} of the map $\varphi_*$
(with respect to $\rho$). For a discussion see
subsection~\ref{subsec:exchange} below, and the
forthcoming~\cite{winter:massar:cp}.

\section{Discussion}
\label{sec:discussion}
We have introduced a separation into extrinsic and intrinsic data
of a quantum measurement. It was shown to have definite minimal
rates for either of these, and that it encompasses all previously
known results on ``meaningful'' data in quantum measurements.
A particular advantage of theorem~\ref{thm:ex-and-in}
before theorems~\ref{thm:massar:popescu}
and~\ref{thm:POVM:compr:1} is that it not even requires a new
POVM (which might be experimentally difficult to realize).
Instead, it can be understood as a mere re--interpretation of the data
delivered by ${\bf a}^{\otimes l}$:
in fact, by our construction in the proof of theorem~\ref{thm:ex-and-in}
for all $\nu$ and $j^l$ either $A^{(\nu)}_{j^l}$ is $0$
or very close to a multiple of $a_{j^l}$, in the sense of (CM). Hence
the random variable $\Nu$, defined as a function of $j^l$:
\begin{equation}
  \label{eq:nu:from:j}
  \Pr\{\Nu=\nu|j^l\}=\frac{x_\nu}{\lambda_{j^l}}\tr\left(\omega A^{(\nu)}_{j^l}\right)
                    =x_\nu\frac{\tr\left(\omega A^{(\nu)}_{j^l}\right)}{\tr(\omega a_{j^l})},
\end{equation}
(up to a scaling factor, close to $1$ for typical $j^l$),
is almost independent from the source
ensemble $\{\sigma_k,q_k\}$ in (CP). More precisely,
$$\sum_{k \nu j^l} \frac{1}{2}
      \left| q_k x_\nu\tr(\sigma_k A^{(\nu)}_{j^l})
            -q_k \tr(\sigma_k a_{j^l})\Pr\{\Nu=\nu|j^l\}\right|\leq \epsilon,$$
and in fact, we even have
$$\sum_{\nu j^l} \frac{1}{2}
      \left\| x_\nu\sqrt{\omega}A^{(\nu)}_{j^l}\sqrt{\omega}
             -\Pr\{\Nu=\nu|j^l\}\sqrt{\omega}a_{j^l}\sqrt{\omega}\right\|_1\leq \epsilon.$$
\par
This means that one can reproduce the statistics of the whole
diagram in figure~\ref{fig:bottle} from the outcomes of
${\bf a}^{\otimes l}$, by \emph{inventing}
the $\nu$ distributed according to eq.~(\ref{eq:nu:from:j}).
This gives a new view on the extrinsic/intrinsic separation:
rather than replacing the original POVM by a fancy construction,
one can from the original data $j^l$ compute the extrinsic
data $\nu$, and conditional on that the intrinsic part.
Then one can sucessfully \emph{pretend} that this separation 
was delivered by the mixture of the POVMs ${\bf A}^{(\nu)}$.

\subsection{Data vs.~Information}
\label{subsec:data-info}
One (as it turns out, rather careless) interpretation of our
result could be that the ``useful'' information produced by
the POVM ${\bf a}$ amounts to $I(\lambda;\hat{\rho})$.
This in itself is not yet precise, so lets fix ``information''
to mean ``communicable information'' in the sense of
Shannon~\cite{shannon:info}: for any source $\{\sigma_i,\mu_i\}$
with average $\sum_i \mu_i\sigma_i=\rho$ the source and
measurement outcome are random variables $X$ and $Y$
with a joint distribution
$$\Pr\{X=i,Y=j\}=\mu_i \tr(\sigma_i a_j),$$
and the mutual information of these is
$$I(X\wedge Y)=H(X)+H(Y)-H(XY).$$
We repeat here the discussion of~\cite{winter:massar:POVMcompr}
regarding the relation between this quantity and $I(\lambda;\hat{\rho})$:
\par
Observe first that the joint distribution of $X$ and $Y$ can
be rewritten as
\begin{equation*}\begin{split}
  \Pr\{X=i,Y=j\} &=\tr\left(\rho^{-1/2}\mu_i\sigma_i\rho^{-1/2}\sqrt{\rho}a_j\sqrt{\rho}\right) \\
                 &=\lambda_j \tr(\hat{\rho}_j S_i),
\end{split}\end{equation*}
where the $S_i=\rho^{-1/2}\mu_i\sigma_i\rho^{-1/2}$ form a POVM
(compare~\cite{hjw:classification} where this correspondence between
POVMs and ensembles was used to classify the latter with given density matrix).
But here the Holevo bound~\cite{holevo:bound}
applies, with the ensemble
$\{\hat{\rho}_j,\lambda_j\}$, and thus we have proved:
\begin{thm}
  \label{thm:infobound}
  Let $\{\sigma_i,\mu_i\}$ be any ensemble whose average state
  $\sum_i \mu_i\sigma_i$ equals $\rho$. Define random
  variables $X,Y$ with joint distribution
  $$\Pr\{X=i,Y=j\}=\mu_i\tr(\sigma_i a_j)$$
  (this is the probability for $\sigma_i$ to occur
  and that $j$ is observed on this state). Then
  $$I(X\wedge Y)\leq I(\lambda;\hat{\rho}).$$
  \qed
\end{thm}
Note that in general maximization over the ensemble $\{\sigma_i,\mu_i\}$
(yielding the \emph{accessible information}
$$J_\rho({\bf a})=I_{\rm acc}(\lambda;\hat{\rho}),$$
because in the above proof it corresponds to an information maximization
over the POVM $S_i$)
does not achieve the upper bound: see~\cite{holevo:bound}, where it is shown that
it does if and only if all the $\hat{\rho}_j$ commute.
\par
Furthermore, by a result from~\cite{holevo:bound:1}
$$J_{\rho^{\otimes l}}({\bf a}^{\otimes l})=l J_\rho({\bf a}),$$
hence the gap remains even asymptotically!
For further discussion of this point we refer the reader
to~\cite{winter:massar:POVMcompr}, section VII\,C. We record
here only the consequence that one ought to distinguish between
\emph{data} (collected by measurement) and \emph{information}
(about a property of the states): the latter is never
larger than the former, and typically in quantum
situations it is strictly less. However, this seems nothing to
worry about: after all, this is an observation quite familiar
from our experience, though it is worth stressing that in
the present context it is a purely quantum phenomenon.
\par
Peter Shor has remarked the notable fact that in the presence of entanglement,
however, this distinction disappears: the entanglement--assisted
capacity~\cite{bennett:shor:ent:ass:cap} for the quantum--classical channel
that is represented by our POVM, i.e. $\varphi_*$ from
eq.~(\ref{eq:POVM:op1}), with the average of the sent symbols
required to be $\rho$ (this means that in the formula for the entanglement--assisted
capacity one has to put a purification of $\rho$)
coincides with our $I(\lambda;\hat\rho)$! In fact, our result
can be understood as a weak version of the conjectured
``Quantum Reverse Shannon Theorem''~\cite{bennett:shor:ent:ass:cap},
for quantum--classical channels.
\par\medskip
To end this part of the discussion note that the bound of
theorem~\ref{thm:ex-and-in} in the case of a maximally refined
measurement is simply the von Neumann entropy $H(\rho)$
of the source, and this regardless of the nature of the
POVM and of the source. In this sense, there is ``democracy
among measurements'', at least the maximally refined ones.
\par
It is thus appealing to view our result as a dual to the 
creation of a density operator by mixing pure states: it is
well known that in any representation $\rho=\sum_i p_i\sigma_i$,
with pure states $\sigma_i$, $H(p)\geq H(\rho)$, with equality
iff the $\sigma_i$ are mutually orthogonal eigenstates of $\rho$:
hence, $H(\rho)$ is the \emph{minimum entropy needed to generate
$\rho$}. In the present work we identify $H(\rho)$ as the
\emph{maximum entropy of measurement data correlated to $\rho$}.

\subsection{Holevo bound}
\label{subsec:holevobound}
Here we show how to turn around the previous argument to
actually \emph{prove} the Holevo information bound.
The statement is as follows:
\begin{thm}[Holevo~\cite{holevo:bound}]
  \label{thm:holevo}
  Let $\{\hat{\rho}_j,\lambda_j\}_{j=1,\ldots,m}$ be an ensemble of states with
  average $\rho=\sum_j \lambda_j\hat{\rho}_j$, and $(S_i)_{i=1,\ldots,n}$ a POVM.
  Define the joint distribution of random variables $Y$, $X$ to be
  \begin{equation}
    \label{eq:YX:distribution}
    \Pr\{Y=j,X=i\}=\lambda_j \tr(\hat{\rho}_j S_i).
  \end{equation}
  Then the inequality
  $$I(Y\wedge X)\leq I(\lambda;\hat{\rho})
                   = H(\rho)-\sum_j \lambda_j H(\hat{\rho}_j)$$
  holds.
\end{thm}
\begin{beweis}
  To begin with, observe that eq.~(\ref{eq:YX:distribution})
  may be rewritten as
  $$\Pr\{Y=j,X=i\}=p_i \tr(\sigma_i a_j),$$
  with $a_j=\rho^{-1/2}\lambda_j\hat{\rho}_j\rho^{-1/2}$
  and the ensemble $\{\sigma_i,p_i\}$, where
  $p_i\sigma_i=\sqrt{\rho}S_i\sqrt{\rho}$.
  Now consider i.i.d.
  realizations $X_1,Y_1,\ldots,X_l,Y_l$ of the pair $X,Y$.
  We shall apply theorem~\ref{thm:ex-and-in} to ${\bf a}^{\otimes l}$
  and $\rho^{\otimes l}$, with parameter $0<\epsilon<1$.
  Hence, for ${\bf A}=\sum_\nu x_\nu {\bf A}^{(\nu)}$ and the
  ensemble $\{\sigma_{i^l},p_{i^l}\}$ the condition (CP)
  holds. Let us define random variables $\xi,\upsilon$ by
  $$\Pr\{\upsilon=j^l,\xi=i^l\}=p_{i^l}\tr(\sigma_{i^l}{\bf A}_{j^l}).$$
  Then we may calculate (with $f(\epsilon):=\epsilon(\log m+2\log n)$)
  \begin{equation*}\begin{split}
    l I(Y\wedge X) &=    I(X^l\wedge Y^l) \\
                   &\leq I(\xi\wedge\upsilon)+lf(\epsilon)+4 \\
                   &\leq I(\xi\wedge\nu\mu)+lf(\epsilon)+4         \\
                   &=    I(\xi\wedge\nu)+I(\xi\wedge\mu|\nu)+lf(\epsilon)+4 \\
                   &\leq \makebox[\lueckeeins]{$0$}
                                        +\makebox[\lueckezwei]{$\log M$}
                                                            +lf(\epsilon)+4 \\
                   &\leq l I(\lambda;\hat{\rho})+O(\sqrt{l})+lf(\epsilon)+4.
  \end{split}\end{equation*}
  Only classical entropy relations have been used: line $2$
  is by lemma~\ref{lemma:H:cont} stated below, line $3$ is
  by data processing, as $\upsilon$ is a function of $\nu$ and $\mu$,
  line $4$ is a standard identity, and line $5$ by
  independence of $\nu$ and $\xi$ and the standard
  inequality $I(\xi\wedge\mu|\nu)\leq H(\mu)$.
  \par
  Now divide by $l$ and let $l\rightarrow\infty$:
  $$I(Y\wedge X)\leq I(\lambda;\hat{\rho})+\epsilon(\log m+2\log n).$$
  As $\epsilon>0$ was arbitary, the theorem follows.
\end{beweis}
\begin{lemma}[Fano~\cite{fano}]
  \label{lemma:H:cont}
  Let $P$ and $Q$ be probability distributions on a set
  with finite cardinality $a$, such that
  $\frac{1}{2}\|P-Q\|_1\leq\lambda$. Then
  $$|H(P)-H(Q)| \leq \lambda\log a+2H(\lambda,1-\lambda).$$
  \qed
\end{lemma}
\par
The reader may want to compare this proof to our earlier one
in~\cite{winter:massar:POVMcompr}: despite similarities they
are conceptually completely different! In fact, there we introduced
the Holevo mutual information as a certain fidelity measure
(which may seem slightly artificial) and applied theorem~\ref{thm:POVM:compr:1},
while here we directly exploit the ``bottleneck'' nature of our main result
(compare again fig.~\ref{fig:bottle}), thus
providing a much more natural approach.

\subsection{Fixed source ensemble and classical case}
\label{subsec:fixedsource:and:classical}
Our approach has concentrated on universal properties of the POVM,
leaving the source as free as possible. What happens if we fix the source
$\{\rho_i,p_i\}$? Note firstly that the whole situation is
fully classical now, as we only have to regard the correlation
between source issues $X=i$ and measurement results $Y=j$.
\par
Thus it is modelled by the classical case of the initial problem:
the source is $\{\ketbra{i},p_i\}$, and the POVM ${\bf b}$
consists of operators
$$b_j=\sum_i \tr(\rho_i a_j) \ketbra{i}.$$
This model has the same joint statistics  of $i$ and $j$
as the above described one (most generally, $b_j$ can be
any operator with eigenbasis $\{\ket{i}\}$).
\par
Now observe the following: as long as the POVMs ${\bf A}^{(\nu)}$ are
diagonal in the basis $\{\ket{i^l}\}$, too (this is the
classicality condition for the POVMs), the validity of (CP)
for all ensembles with average
$$P=\sum_i p_i\ketbra{i}$$
is implied by its validity for the ensemble $\{\ketbra{i},p_i\}$.
This is because source states $\rho_{i'}$ and $\sum_i \ketbra{i}\rho_{i'}\ketbra{i}$
produce the same statistics, so only sources consisting of mixtures
of the $\ketbra{i}$ have to be considered. The condition (CP) for
them clearly is implied by its validity for $\{\ketbra{i},p_i\}$.
\par
At this point theorems~\ref{thm:ex-and-in} and~\ref{thm:strong:converse}
can be applied: because the induced ensemble for source state $P$
and POVM ${\bf b}$ is $\{\hat{\sigma}_j,\lambda_j\}$, with
\begin{align*}
  \lambda_j      &=\sum_i p_i\tr(\rho_i a_j)=\tr(\rho a_j), \\
  \hat{\sigma}_j &=\sum_i \frac{1}{\lambda_j}p_i\tr(\rho_i a_j)\ketbra{i},
\end{align*}
we obtain $I(X\wedge Y)$, that is the Shannon mutual information between
the source and the measurement, as the rate of intrinsic data.
More precisely, we can perform a data separation by postprocessing,
according to the prescription of the beginning of this section,
eq.~(\ref{eq:nu:from:j}), into extrinsic $\nu$, almost inpendent of $i^l$,
and intrinsic $j^l$ depending on $i^l$ and $\nu$.
\par
However, this is not exactly what we set out to initially:
theorem~\ref{thm:ex-and-in} allows us to
decompose the $b_{j^l}$ into convex combinations of operators
$$B^{(\nu)}_{j^l}=\sum_{i^l} \beta^{(\nu)}_{j^l|i^l}\ketbra{i^l},$$
but it is not clear that these can be obtained from POVMs ${\bf A}^{(\nu)}$,
in the sense that
$$\forall\nu\forall j^l\forall i^l\quad
      \beta^{(\nu)}_{j^l|i^l}=\tr\left(\rho_{i^l} A^{(\nu)}_{j^l}\right).$$
For this to hold the vectors $\left(\beta^{(\nu)}_{j^l|i^l}\right)_{i^l}$
(for all $j^l$) must belong to the cone spanned by the vectors
$(\bra{\psi}\rho_{i^l}\ket{\psi})_{i^l}$.
It is conceiveable that under this condition the obtainable intrinsic data
rate increases. We have to leave this interesting question for the moment.
\par
For classical sources and measurements we thus obtain that intrinsic data
equals mutual information. On the other hand, we can come back to
their being distinct in truly quantum situations: we pointed out
in subsection~\ref{subsec:data-info} that the maximum of $I(X\wedge Y)$
over all sources with average $\rho$ gives the accessible information
$I_{\rm acc}(\lambda;\hat\rho)$ of the ensemble
$\{\hat{\rho}_j,\lambda_j\}$, which in general is less than $I(\lambda;\hat\rho)$.
The difference can be accounted for by considering that the sources in
this maximization are of the special i.i.d.~type (on $l$--blocks),
while (CM) implies (CP) even for sources of entangled states, as
long as their average is $\omega=\rho^{\otimes l}$.
This should be viewed especially in the light of the conjecture implied
in subsection~\ref{subsec:AVS}.

\subsection{Sufficient statistics}
\label{subsec:sufficient}
The reader familiar with classical statistical theories may have
been reminded by our above discussion of the concept of
\emph{sufficient statistics}, at least when the quantum
source and the observation are essentially classical, i.e. when
all the $\rho_i$ and $a_j$ commute: the former are then just
probability distributions and the latter form a statistical decision
rule, with distribution of $j$ conditional on $i$ denoted $q(j|i)$.
As there is also a distribution $p_i$ on the $i$ we have here
a statistical model in the sense of estimation theory (we refer
the reader to~\cite{lehmann:casella} for detailed explanations).
\par
We will consider the values of $i$ and $j$ as random variables:
then a \emph{sufficient statistics} is a random variable $k$ which
is a function of $j$ (whose distribution conditional on $i$ we
denote $\tilde{q}(k|i)$),
such that the distribution of $j$ conditional
on $k$ is independent of $i$:
$$\Pr\{j|k\}=\Pr\{j|k,i\}\qquad \forall i.$$
Let us denote these conditional probabilities by $r(j|k)$.
\par
This implies that we can simulate the distribution of $j$ conditional
on $i$ from $k$:
$$q(j|i)=\sum_k r(j|k)\tilde{q}(k|i).$$
In words, to each entry $k$ of the new data record there
exists a distribution on the $j$ of the original data record
such that the latter's distribution is recovered
as a convolution; in terms of stochastic maps
$q$ is factorized into $\tilde{q}$ and $r$:
$$i \stackrel{\tilde{q}}{\longrightarrow} k \stackrel{r}{\longrightarrow} j.$$
\par
On the other hand, our theorem~\ref{thm:ex-and-in} provides
something appearing to be dual to this (apart from holding only
approximately and in an asymptotic setting; these things
are easily introduced in sufficient statistics, too):
a random varible $\nu$ with distribution $x$,
independent of $i$ and $j$, and conditional on it
a stochastic map $a_\nu(j|i)$ such that
$$q(j|i)=\sum_\nu x_\nu a_\nu(j|i).$$
In a diagram:
\begin{align*}
    &\makebox[\lueckedrei]{$\,\,\nu$}           \\
    &\makebox[\lueckedrei]{$\,\,\downarrow$}    \\
  i &\overbrace{\stackrel{{\widetilde{Q}_\nu}}{\longrightarrow} \mu \stackrel{R_\nu}{\longrightarrow}} j.
\end{align*}
Like $k$ in the case of sufficient statistics, the pair
$\mu\nu$ is a function of $j$, but unlike there, where $\tilde{q}$
and $r$ were stochastic maps with independent sources of
randomness (when stochastic maps are viewed as set function valued
random variables, this is expressed by the independence of $\tilde{q}$
and $r$), the maps $\widetilde{Q}$ and $R$ draw their
randomness from the same source $\nu$.
\par
In summary, there is no direct isomorphism between our
concept of data reduction and sufficient statistics (which,
too, can be used to reduce the entropy of data sets): the latter
appears as a special case where the maps $\widetilde{Q}$
and $R$ are independent.

\subsection{Entropy exchange}
\label{subsec:exchange}
We want to discuss an application of theorem~\ref{thm:cp-map}
to the \emph{entropy exchange} of quantum operations, introduced
by Schumacher~\cite{schumacher:exchange} (and previously
by Lindblad~\cite{lindblad:exchange}):
for a quantum operation $\varphi_*$ in the form~(\ref{eq:kraus})
it is defined as
$$S_e(\rho;\varphi_*)=H(W),\text{ with }W_{jk}=\tr(V_j\rho V_k^*).$$
It can be shown to be independent of the Kraus representation,
by identifying it with the entropy increase in an initially
pure environment of the system by a Stinespring dilation of
$\varphi_*$, see~\cite{schumacher:exchange}. In the latter
work a number of interesting relations between $S_e$ and
other entropic quantities are shown.
\par
In particular, returning to the notation of section~\ref{sec:other:appl},
it is shown that there is a (in this sense, minimal) Kraus
representation of $\varphi_*$ such that $H(\lambda)=S_e(\rho;\varphi)$.
Because of $I(\lambda;\hat{\rho})\leq H(\lambda)$ (this is
simply data processing inequality~\cite{ahlswede:loeber}),
we conclude
$$\Sigma(\rho;\varphi_*)\leq S_e(\rho;\varphi_*).$$
\par
By the derivation this quantity may be dubbed \emph{genuinely quantum
entropy exchange} of a channel, as it is that part of the noise
that cannot be accounted for classically.
\par
From a different point of view, in fact also the \emph{maximum}
of $I(\lambda;\hat{\rho})$ over all Kraus representations of
$\varphi_*$ (compare eq.~(\ref{eq:sigma})) is interesting: in
a cryptographic setting, where $\varphi_*$ connects users A and B,
and is controlled by an eavesdropper E, it is the amount of
data collected by E about A's messages in the worst case.
\par
A deeper investigation of these concepts is relegated to another
occasion~\cite{winter:massar:cp}.
%
%

\subsection{An open problem}
\label{subsec:AVS}
An interesting and challenging question is about the amount of data
collected by ${\bf a}$ under the hypothesis of an \emph{arbitrarily varying
source (AVS)}, instead of the i.i.d. model considered here:
\par
An AVS is a collection of source ensembles $\{\rho_{is},p_{is}\}$
(with average state $\rho_s$),
labelled by $s\in\fset{S}$, which we make into a discrete
memoryless source by considering the ensembles
(labelled by $s^l\in\fset{S}^l$)
$$\{\rho_{i^ls^l},p_{i^ls^l}\}_{i^l}.$$
The idea is that at each position $k=1,\ldots,l$ the source
may be arbitrarily in one of the \emph{internal states} $s\in\fset{S}$.
We have no --- not even statistical information --- about $s$, so
our data separation must work for \emph{all} $s^l\in\fset{S}^l$:
formally the condition on ${\bf A}=\sum_\nu x_\nu {\bf A}^{(\nu)}$ is
\begin{equation}
  \forall s^l\quad \sum_{j^l} \frac{1}{2}
       \left\|\sqrt{\omega(s^l)}(a_{j^l}-A_{j^l})\sqrt{\omega(s^l)}\right\|_1
                                                         \leq \epsilon,\tag{AVCM}
\end{equation}
where $\omega(s^l)=\rho_{s_1}\otimes\cdots\otimes\rho_{s_l}$ is
the average state of the source when in internal state $s^l$.
\par
A natural candidate for the minimum data rate of the ${\bf A}^{(\nu)}$
seems to be
$$\max \bigl\{I(\lambda;\hat{\rho}):\rho\in\overline{\conv}\{\rho_s:s\in\fset{S}\}
       \bigr\},$$
with $\lambda_j\hat{\rho}_j=\sqrt{\rho}a_j\sqrt{\rho}$, and $\overline{\conv}$
denoting the closed convex hull.
\par
If this is true, then in particular the quantity
$$\Delta({\bf a})=\max_\rho I(\lambda;\hat{\rho})$$
is \emph{the} amount of data collected by ${\bf a}$, regardless
of any source ensemble.
%
%

\begin{acknowledgments}
  I am indebted to Serge Massar for his introducing me to the
  problem addressed in this paper and for interesting discussions,
  and to Hiroshi Nagaoka for pointing me at the
  possible relation between the present approach and
  sufficient statistics. Thanks to Peter Shor who supplied the insight
  that the difference between data and information disappears in
  the presence of entanglement.
  I thank Masanao Ozawa for pointing out to me that
  theorem~\ref{thm:cp-map}, initially only formulated for operations,
  is in fact valid for instruments. Part of this work was done
  during my stay at the ERATO project ``Quantum Computation and
  Information'', Tokyo (August/September 2001). I thank the members
  of the project for their hospitality, and especially Keiji Matsumoto
  for discussions on the content of the appendix, on which I also
  enjoyed conversation with Richard Jozsa and Masahide Sasaki.
  Last but not least, special thanks are due to Marco P.~Carota for
  constant encouragement during the course of this work.
\end{acknowledgments}

\appendix
\section{Canonical purifications}
\label{sec:app:fidelity}
In this appendix we collect a few facts about mixed state fidelity and
a certain kind of purification of mixed states, which we call \emph{canonical},
that seem not to be widely known.
These are used in the main text, but seem to be of interest in their own
right.
\par\medskip
For the state $\omega$ on ${\cal H}_1$ consider
a purification $\ket{\psi}=\sum_i \sqrt{r_i}\ket{i}\otimes\ket{i}$
on a bipartite system ${\cal H}_1\otimes{\cal H}_2$,
that we already have put in Schmidt polar form.
Then on both systems there exist ($\R$--linear) complex conjugation
maps with respect to the basis $\{\ket{i}\}$:
$$\ket{\phi}=\sum_i \alpha_i\ket{i}\longmapsto 
                        \sum_i \overline{\alpha_i}\ket{i}=:\ket{\overline{\phi}}.$$
Then, with $\ket{I}=\sum_i \ket{i}\otimes\ket{i}$, it can be checked that
\begin{equation*}\begin{split}
  \ketbra{\psi} &= (\sqrt{\omega}\otimes\1)\ketbra{I}(\sqrt{\omega}\otimes\1) \\
                &= (\1\otimes\sqrt{\omega})\ketbra{I}(\1\otimes\sqrt{\omega}),
\end{split}\end{equation*}
see also the following lemma~\ref{lemma:canonical:purifications}. Then
\begin{equation*}\begin{split}
  &\left(\1\otimes\sqrt{S_k}\right)\!\ketbra{\psi}\!\left(\1\otimes\sqrt{S_k}\right)           \\
  &\phantom{===}
   = \left(\1\otimes\sqrt{S_k}\sqrt{\omega}\right)\!
                                     \ketbra{I}\!\left(\1\otimes\sqrt{S_k}\sqrt{\omega}\right) \\
  &\phantom{===}
   =q_k(\1\otimes U_k)\bigl[
                         (\1\otimes\sqrt{\tau_k})\ketbra{I}(\1\otimes\sqrt{\tau_k})
                      \bigr](\1\otimes U_k^*)                                                  \\
  &\phantom{===}
   =q_k(\1\otimes U_k)\ketbra{t_k}(\1\otimes U_k^*),
\end{split}\end{equation*}
the third line introducing $q_k\tau_k=\sqrt{\omega}S_k\sqrt{\omega}$ on ${\cal H}_2$,
and the polar decomposition $\sqrt{S_k}\sqrt{\omega}=U_k\sqrt{q_k\tau_k}$, the fourth
the \emph{canonical purification} $\ket{t_k}$ on ${\cal H}_1\otimes{\cal H}_2$ of
$\tau_k$ (with respect to $\ketbra{I}$), see lemma~\ref{lemma:canonical:purifications} below.
By this lemma we can infer
\begin{equation*}\begin{split}
  \tr_{{\cal H}_2}\bigl[\ketbra{\psi}(\1\otimes S_k)\bigr]
                                  &= q_k \tr_{{\cal H}_2} \ketbra{t_k} \\
                                  &= q_k\overline{\tau_k},
\end{split}\end{equation*}
with the complex conjugated operator $\overline{\tau_k}$, which is defined as
$$\overline{\tau_k}=\sum_i \ketbra{\overline{\phi_i}},\text{ if }
            \tau_k =\sum_i \ketbra{\phi_i}.$$
Note that this is uniquely defined, regardless of
the convex decomposition chosen, and in particular independent of
the phases of the $\ket{\phi_i}$.
\par
The ensemble $\{\overline{\tau_k},q_k\}$ has average $\overline{\omega}=\omega$,
and conversely, the above
formulas show how to induce any ensemble
$\{\sigma_k,q_k\}$ for $\omega$ on ${\cal H}_1$: let
$S_k=\omega^{-1/2} q_k\overline{\sigma_k} \omega^{-1/2}$
(this was noted before in~\cite{hjw:classification} in the context of classifying
ensembles with a given density operator).
\begin{lemma}[``Pretty good purifications'']
  \label{lemma:canonical:purifications}
  Consider orthonormal bases of spaces ${\cal H}_1$
  and ${\cal H}_2$, both denoted $\{\ket{i}\}$, and introduce
  $\ket{I}=\sum_i \ket{i}\otimes\ket{i}$. As before, we denote the
  complex conjugation with respect to this basis by $\overline{\phantom{x}}$.
  Then for a state $\rho=\sum_i \alpha_i\ketbra{\psi_i}$ (in diagonalized form),
  \begin{align*}
    \ketbra{r} &=\left(\sqrt{\rho}\otimes\1\right)\ketbra{I}\left(\sqrt{\rho}\otimes\1\right),\text{ with} \\
       \ket{r} &=\sum_i \sqrt{\alpha_i}\ket{\psi_i}\otimes\ket{\overline{\psi_i}},
  \end{align*}
  is a purification of $\rho$. We call it the \emph{canonical purification}
  with respect to $\ket{I}$. (Note that this definition makes sense as it is
  independent of phases in the $\ket{\psi_i}$).
  \par
  If $\ketbra{s}$ is the canonical purification of another state $\sigma$
  then for the fidelity between these:
  \begin{equation}
    \label{eq:canonical:fid}
    F(\ket{r},\ket{s})=|\langle r\ket{s}|^2=\left(\tr\sqrt{\rho}\sqrt{\sigma}\right)^2.
  \end{equation}
  Furthermore
  \begin{align}
    \label{eq:canonical:vs:optimal}
    \tr\sqrt{\rho}\sqrt{\sigma} &\geq 1-\sqrt{\|\rho-\sigma\|_1},    \\
    \label{eq:canonical:tracenorm}
    \frac{1}{2}\bigl\| \ketbra{r}-\ketbra{s} \bigr\|_1 &\leq \sqrt[4]{4 \|\rho-\sigma\|_1}.
  \end{align}
\end{lemma}
\begin{beweis}
  The formula for the canonical purification is a straightforward calculation.
  With its help, it is also straightforward to check the fidelity
  identity, eq.~(\ref{eq:canonical:fid}).
  \par
  Now for the last two estimates: begin with
  \begin{equation*}\begin{split}
    1-\tr\sqrt{\rho}\sqrt{\sigma} &=\tr\left(\sqrt{\rho}\bigl(\sqrt{\rho}-\sqrt{\sigma}\bigr)\right)  \\
                           &\leq \left\| \sqrt{\rho}(\sqrt{\rho}-\sqrt{\sigma}) \right\|_1         \\
                           &\leq \left\| \sqrt{\rho} \right\|_2
                                 \left\| \sqrt{\rho}-\sqrt{\sigma} \right\|_2                  \\
                           &\leq \left\| \sqrt{\left| \rho-\sigma \right|} \right\|_2       \\
                           &=    \sqrt{\|\rho-\sigma\|_1},
  \end{split}\end{equation*}
  invoking two nontrivial inequalities: in the third line we
  use Cor.~IV.2.6 of~\cite{bhatia:matrix} (which is a kind of H\"older or
  Cauchy--Schwarz inequality), in the fourth line Thm.~X.1.3 from the same book.
  \par
  Finally, use the well known identity
  $$\frac{1}{2}\bigl\| \ketbra{r}-\ketbra{s} \bigr\|_1=\sqrt{1-F(\ket{r},\ket{s})}$$
  to obtain
  \begin{equation*}\begin{split}
    \frac{1}{2}\bigl\| \ketbra{r}-\ketbra{s} \bigr\|_1
                                    &=    \sqrt{1-|\langle r\ket{s}|^2}                \\
                                    &\leq \sqrt{2}\sqrt{1-|\langle r\ket{s}|}          \\
                                    &=    \sqrt{2}\sqrt{1-\tr\sqrt{\rho}\sqrt{\sigma}} \\
                                    &\leq \sqrt{2}\sqrt[4]{\|\rho-\sigma\|_1},
  \end{split}\end{equation*}
  which we wanted to show.
\end{beweis}
\begin{rem}
  \label{rem:fidelity}
  Observe $\tr\sqrt{\rho}\sqrt{\sigma}\leq \|\sqrt{\rho}\sqrt{\sigma}\|_1$,
  the square of this latter quantity being
  known as the \emph{(mixed state) fidelity}~\cite{jozsa:msf}.
  By theorems by Uhlmann~\cite{uhlmann:fidelity} and Jozsa~\cite{jozsa:msf}
  the mixed state fidelity $F(\rho,\sigma)=\|\sqrt{\rho}\sqrt{\sigma}\|_1^2$
  equals the maximum over the pure state fidelities of all possible
  purifications of $\rho$ and $\sigma$.
  Because of well known relations between mixed state fidelity and trace norm
  distance (see~\cite{fuchs:vandegraaf}), more precisely
  \begin{equation}
    \label{eq:fidelity:tracenorm}
    1-\sqrt{F(\rho,\sigma)} \leq \frac{1}{2}\|\rho-\sigma\|_1
                            \leq \sqrt{1-F(\rho,\sigma)},
  \end{equation}
  the lemma tells us that at least for (mixed state) fidelity close to $1$
  the canonical purifications are not too far off the optimum with respect
  to (pure state) fidelity.
\end{rem}

\section{Typical sequences and subspaces}
\label{sec:typical}
For a probability distribution $P$ on the finite set $\fset{X}$
define set of \emph{typical sequences} (with $\delta>0$)
$$\fset{T}^l_{P,\delta}=\left\{x^l:\forall x\ |N(x|x^l)-lP_x|\leq
                               \delta\sqrt{l}\sqrt{P_x(1-P_x)}\right\},$$
where $N(x|x^l)$ counts the number of occurences of
$x$ in the word $x^l=x_1\ldots x_n$.
\par
For a state $\rho$ fix eigenstates $e_1,\ldots,e_d$ (with
eigenvalues $R_1,\ldots,R_d$) and
define for $\delta>0$ the \emph{typical projector} as
$$\Pi^l_{\rho,\delta}=\sum_{t^l\in\fset{T}^l_{R,\delta}}
                              e_{t_1}\otimes\cdots\otimes e_{t_l}.$$
For a collection of states $\hat{\rho}_j$, $j=1,\ldots,m$, and
$j^l\in[m]^l$ define the \emph{conditional typical projector} as
$$\Pi^l_{\hat{\rho},\delta}(j^l)=\bigotimes_j 
                              \Pi^{I_j}_{\hat{\rho}_j,\delta},$$
where $I_j=\{k:j_k=j\}$ and $\Pi^{I_j}_{\hat{\rho}_j,\delta}$
is meant to denote the typical projector of the state $\hat{\rho}_j$
on the subsystem composed of the tensor fatcors $I_j$ in the tensor
product of $l$ factors.
From~\cite{winter:ieee_strong} we cite the following 
properties of these projectors:
\begin{align}
  \label{eq:typical:prob}
  \tr(\rho^{\otimes l}\Pi^l_{\rho,\delta})            &\geq 1-\frac{d}{\delta^2}, \\
  \label{eq:cond:typical:prob}
  \tr(\hat{\rho}_{j^l}\Pi^l_{\hat{\rho},\delta}(j^l)) &\geq 1-\frac{md}{\delta^2}, \\
  \label{eq:typical:prob:general}
  \tr(\hat{\rho}_{j^l}\Pi^l_{\rho,\delta})            &\geq 1-\frac{m^2d}{\delta^2},
\end{align}
\begin{align}
  \label{eq:typical:upper}
    \tr\Pi^l_{\rho,\delta}            &\leq \exp\left(lH(\rho)+Kd\delta\sqrt{l}\right), \\
  \label{eq:typical:lower}
    \tr\Pi^l_{\rho,\delta}            &\geq \left(1-\frac{d}{\delta^2}\right)
                                            \exp\left(lH(\rho)-Kd\delta\sqrt{l}\right),
\end{align}
\begin{align}
  \label{eq:cond:typical:upper}
    \tr\Pi^l_{\hat{\rho},\delta}(j^l) &\leq \exp\left(lH(\hat{\rho}|P_{j^l})
                                                             +Kmd\delta\sqrt{l}\right), \\
  \label{eq:cond:typical:lower}
    \tr\Pi^l_{\hat{\rho},\delta}(j^l) &\geq \left(1-\frac{md}{\delta^2}\right)
                                            \exp\left(lH(\hat{\rho}|P_{j^l})
                                                             +Kmd\delta\sqrt{l}\right),
\end{align}
for an absolute constant $K>0$, and the empirical distribution
$P_{j^l}$ of letters $j$ in the word $j^l$:
$$P_{j^l}(j)=\frac{1}{l}N(j|j^l).$$
Finally, with
\begin{align*}
  \alpha  &=\exp\left(-lH(\rho)-Kd\delta\sqrt{l}\right), \\
  \alpha' &=\exp\left(-lH(\rho)+Kd\delta\sqrt{l}\right),
\end{align*}
\begin{align*}
  \beta   &=\exp\left(-lH(\hat{\rho}|P_{j^l})+Kmd\delta\sqrt{l}\right), \\
  \beta'  &=\exp\left(-lH(\hat{\rho}|P_{j^l})-Kmd\delta\sqrt{l}\right),
\end{align*}
we have
\begin{equation}
  \label{eq:typical:rho:lower}
    \alpha'\Pi^l_{\rho,\delta}
       \geq \Pi^l_{\rho,\delta}\rho^{\otimes l}\Pi^l_{\rho,\delta}
                                          \geq \alpha\Pi^l_{\rho,\delta},
\end{equation}
\begin{equation}
  \label{eq:cond:typical:rho:upper}
    \beta'\Pi^l_{\hat{\rho},\delta}(j^l)
       \leq \Pi^l_{\hat{\rho},\delta}(j^l)\hat{\rho}_{j^l}\Pi^l_{\hat{\rho},\delta}(j^l)
                                          \leq \beta\Pi^l_{\hat{\rho},\delta}(j^l).
\end{equation}



\end{document}